\documentclass[aps,prd,showpacs,preprintnumbers,amsmath,amssymb,nofootinbib]{revtex4-1}

\usepackage[english]{babel}
\usepackage[utf8]{inputenc} 

\usepackage{dcolumn}
\usepackage{bm}
\usepackage{dsfont}
\usepackage{latexsym,fancyhdr}
\usepackage{graphicx,amsmath} 

\newcommand{\Od}{{\cal O}}

\newcommand{\qint}{\int \frac{d^3 \vec{q}}{(2\pi)^3}}
\newcommand{\tr}{\mbox{tr}}

\newcommand{\im}{\mbox{Im}\,}
\newcommand{\re}{\mbox{Re}\,}
\newcommand{\sgn}{\mbox{sgn}}
\newcommand{\diag}{\mbox{diag}}

\newcommand{\modq}{\vert \vec{q} \vert}
\newcommand{\modp}{\vert \vec{p} \vert}

\newcommand{\mpichsq}{M^2_{\pi^\pm}}
\newcommand{\mpineusq}{M^2_{\pi^0}}
\newcommand{\mpichsqzero}{\hat M^2_{\pi^\pm}}
\newcommand{\mpineusqzero}{\hat M^2_{\pi^0}}
\newcommand{\I}{\mathds{1}}
\newcommand{\condtwo}{\langle \bar q q \rangle}

\newcommand{\tintq}{T\sum_n \int\frac{d^3 \vec{q}}{(2\pi)^3}}

\newcommand{\be}{\begin{equation}}
\newcommand{\ee}{\end{equation}}
\newcommand{\ba}{\begin{eqnarray}}
\newcommand{\ea}{\end{eqnarray}}

\newcommand{\IZ}{{\Bbb Z}}
\newcommand{\IR}{{\Bbb R}}

\newcommand{\gsim}{\raise.3ex\hbox{$>$\kern-.75em\lower1ex\hbox{$\sim$}}}
\newcommand{\lsim}{\raise.3ex\hbox{$<$\kern-.75em\lower1ex\hbox{$\sim$}}}

\begin{document}

\title{Electromagnetic effects in the pion dispersion relation at finite temperature}
\author{A.~G\'omez Nicola}
\email{gomez@fis.ucm.es} \affiliation{Departamento de F\'{\i}sica
Te\'orica II. Univ. Complutense. 28040 Madrid. Spain.}
\author{R.~Torres Andr\'es}
\email{rtandres@fis.ucm.es} \affiliation{Departamento de
F\'{\i}sica Te\'orica II. Univ. Complutense. 28040 Madrid. Spain.}

\begin{abstract}

We investigate the charged-neutral difference in the pion self-energy at finite temperature $T$. Within  Chiral Perturbation Theory (ChPT) we extend a previous analysis performed in the chiral and soft pion limits. Our analysis with physical pion masses leads to additional non-negligible contributions for temperatures typical of a meson gas, including  a momentum-dependent function for the self-energy. In addition, a nonzero imaginary part arises to leading order, which we define consistently in the Coulomb gauge and comes from an infrared enhanced contribution due to thermal bath photons. For distributions typical of a heavy-ion meson gas,  the  charged and neutral pion masses and  their difference depend on temperature through  slowly increasing functions. Chiral symmetry restoration turns out to be ultimately responsible for keeping the charged-neutral mass difference smooth  and compatible with the observed charged and neutral  pion spectra. We study also  phenomenological effects related to the thermal   electromagnetic damping, which gives rise to  corrections for transport coefficients and distinguishes between neutral and charged mean free times.  An important part of the analysis is the  connection with chiral symmetry restoration through the relation of the pion mass difference with the vector-axial spectral function difference, which holds at $T=0$ due to a sum rule in the chiral and soft pion limits. We analyze the modifications of that sum rule including nonzero pion masses and temperature, up to $\Od(T^2)\sim\Od(M_\pi^2)$. Both effects produce  terms making the pion mass difference grow against  chiral-restoring decreasing contributions.  Finally, we analyze the  corrections to the previous ChPT and sum rule results within the resonance saturation framework at finite temperature, including explicitly $\rho$ and $a_1$  exchanges. Our results show that the ChPT result is robust at low and intermediate temperatures, the leading resonance corrections within this framework being $\Od(T^2 M_\pi^2/M_R^2)$ with $M_R$ the involved resonance masses.

\end{abstract}


\maketitle

\section{Introduction and Motivation}
\label{sec:intro}

The study of hadronic properties at finite temperature $T$ is one of the theoretical ingredients needed to understand the behaviour of  matter created in Relativistic Heavy Ion Collision experiments, such as those  in RHIC and LHC (ALICE), as it expands from the onset of local equilibrium to the final freeze-out regime \cite{galekapustabook,qm12proc}. This is especially relevant for chiral symmetry restoration and deconfinement, for which the lattice groups have explored exhaustively the phase diagram and other thermodynamical properties \cite{Aoki:2009sc,Cheng:2009zi,Borsanyi:2010bp,Cheng:2010fe,Bazavov:2011nk}. For the case of vanishing baryon chemical potential, the QCD transition becomes a crossover for the physical case of 2+1 flavours, which makes it especially important to define observables which would behave as order parameters, since different quantities would point to different critical temperatures. Thus, the critical range from  the latest lattice simulations lies within $T_c\sim$ 150-170 MeV.

Several hadron gas features have been studied in different approximations. The Hadron Resonance Gas (HRG) describes the system through the statistical ensemble of all free states thermally available and provides a good description both of lattice thermodynamical data and of experimental hadron yields, when some corrections due to interactions and lattice masses are accounted for \cite{Andronic:2008gu,Huovinen:2009yb}. On the other hand, effective chiral models including explicitly  vector and axial-vector resonances have been successfully used to describe several hadron thermal properties relevant for observables such as the dilepton and photon spectra and $\rho-a_1$ mixing/degeneration at the chiral transition
 \cite{Song:1995ga,Song:1996dg,Rapp:1999ej,Turbide:2003si}.

A systematic and model-independent framework to take into account the relevant light meson degrees of freedom and their interactions is Chiral perturbation Theory (ChPT) \cite{Gasser:1983yg, Gasser:1984gg}. The effective ChPT lagrangian is constructed as an expansion of the form ${\cal L}={\cal L}_{p^2}+{\cal L}_{p^4}+\dots$ where $p$ denotes a meson energy scale compared to the chiral scale $\Lambda_{\chi}\sim$ 1 GeV. Pions are actually the more copiously produced particles after a Heavy Ion Collision and their properties from hadronization to thermal freeze-out can be reasonably described within ChPT. The temperatures involved in that regime are not far from the ChPT applicability range and ChPT has the added value of providing model-independent results. Thus, the meson gas description based on ChPT reproduces fairly well the main qualitative features of the system, such as the chiral restoring behaviour given by the quark condensate \cite{Gerber:1988tt}.
 The introduction of realistic (unitarized) pion interactions improves ChPT, providing a more accurate description of several effects of interest in a Heavy-Ion environment, such as thermal resonances and transport coefficients \cite{Dobado:2002xf,FernandezFraile:2005ka,FernandezFraile:2009mi,FernandezFraile:2008vu,Dobado:2008vt}. This approach has also given rise to a deeper understanding of the scalar-pseudoscalar degeneration pattern taking place at chiral restoration, in agreement with lattice data for meson masses and susceptibilities  \cite{Nicola:2013vma}.  In addition, the virial expansion approach within ChPT, including unitarized corrections, allows to parametrize consistently the deviations from the HRG-like free gas contributions \cite{Dobado:1998tv,Pelaez:2002xf,GomezNicola:2012uc}.

The modification of the pion dispersion relation in the thermal bath has been also analyzed within ChPT. Perturbatively, to one loop the only modification is a shift in the pion mass coming from a tadpole diagram, softly increasing with $T$ \cite{Gasser:1986vb}. At two loops, pions develop a more complicated dispersion relation \cite{Schenk:1993ru} including an absorptive imaginary part, which defines a mean collision rate \cite{Goity:1989gs} responsible for the thermalization mean time and  the mean free path of pions in the thermal bath. This rate is also essential to describe correctly the transport coefficients of the pion gas \cite{FernandezFraile:2009mi}. Corrections to the dispersion relation due to nonzero pion chemical potential during the chemical  nonequilibrium phase have also been studied \cite{FernandezFraile:2009kt}.

In this work we will continue with this program  by studying the modifications of the pion dispersion relation due to electromagnetic (EM) isospin-breaking corrections, including  virtual photon exchange, during the hadronic phase at finite temperature. We will  work within ChPT but corrections due to resonance exchange will also be considered,  within the framework of a sum rule connecting the self-energy difference with vector and axial spectral functions to evaluate the possible impact on chiral symmetry restoration, and explicitly in a resonance saturation model to estimate the range of validity of the ChPT analysis. Electromagnetic corrections are the main source of the charged-neutral mass (or more general, the self-energy) difference and can be consistently studied within ChPT by introducing the relevant lagrangian terms of orders ${\cal L}_{e^2}$, ${\cal L}_{e^2p^2}$ and so on, with $e$ the electric charge considered formally in the chiral expansion as $e^2=\Od (p^2/F^2)$, with $F$ the pion decay constant in the chiral limit.

Our analysis extends, on the one hand, the previously mentioned ChPT studies on the thermal pion dispersion relation and, on the other hand,  previous partial analysis of the isospin breaking of such relation, namely, in the chiral and soft pion limit \cite{Manuel:1998sy,Kapusta:1984gj} and using a Cottingham-like approach within resonance exchange in \cite{Ladisa:1999wx}. We will consider physical pion masses, which will give rise to new effects such as the momentum dependence of the self-energy (a pure thermal effect) and a nonzero imaginary part. In addition, the departure from soft-pion sum rules will complicate the connection with spectral functions. Our analysis provides more realistic results regarding heavy-ion and lattice phenomenology, since the chiral limit is intended to be valid only for temperatures $T\gg M_\pi$, which are not reached in the hadron gas.  In addition, our ChPT analysis will ensure the model independency of the results at low and moderate temperatures taking into account all relevant thermal contributions, which is a benchmark when comparing to resonance exchange models. Besides, as we will explain here, the ChPT leading correction includes certain tadpole-like terms which are not present in the leading resonance saturation diagrams and play an important role at the temperatures considered.  We will concentrate first on the corrections to the real part of the self-energy, including its momentum dependence, but we will see that the nonzero pion mass also induces imaginary parts coming from Landau pure thermal cuts of diagrams both with photon and resonance exchange, the latter remaining as a subleading contribution.  Our present work complements and extends also our previous studies of isospin-breaking corrections in the meson gas \cite{Nicola:2011gq, TorresAndres:2011wd}.

Let us discuss some additional motivations to perform this analysis.

The spectral properties of the particles which constitute the thermal bath are in principle subject to modifications with respect to the vacuum, due to their mutual interactions. These modifications might lead to important observable effects, as it is indeed the case  with the $\rho$(770) meson and its influence in the dilepton spectrum\cite{Song:1995ga,Song:1996dg,Rapp:1999ej,Dobado:2002xf}. However,  the temperature dependence of the masses of pions and other light mesons is usually not included in phenomenological analysis of hadron yields \cite{Andronic:2008gu} despite the fact that the dispersion relation enters directly in the particle number distribution. In addition, the very same expansion dynamics is also in principle influenced by the thermal change in the pion dispersion relation.

The importance of the pion dispersion relation in the pressure and equation of state  and thus in the hadron gas expansion has been discussed in  \cite{Rapp:1995py}. On the other hand, a detailed analysis of the impact of the thermal pion mass shift in freeze-out parameters \cite{Zschocke:2009wt} shows a tiny effect from  $M_\pi (T)$, taken as that predicted by one-loop ChPT and hence very soft and increasing. The reason is that at low temperatures the shift is negligible while at higher temperatures, when it becomes sizable, pion momenta are distributed near $p\sim T$ so that the mass terms become small in the dispersion relation. In \cite{Zschocke:2009wt} it is also pointed out that an increasing temperature-dependent pion mass is consistent with the existence of hadron-like states prior to hadronization, with a mass larger than their vacuum value, which could explain the experimentally observed quark number scaling in elliptic flow.

What we intend to address here in this phenomenological context is, first, how EM corrections modify the prediction of a slowly increasing pion mass, at leading order in ChPT. In addition, we want to examine possible sizable differences between neutral and charged self-energies with temperature and momentum, which could be of phenomenological interest when comparing charged and neutral pion distributions.

Neutral pion distributions have been measured  experimentally in recent Heavy Ion Collisions experiments at RHIC in PHENIX \cite{Adcox:2001jp,Adler:2003qi} and STAR \cite{Abelev:2009wx} as well as in more recent  ALICE (LHC) measurements \cite{ConesaBalbastre:2011zi,Kharlov:2012na}. The comparison between neutral and charged pion spectra for STAR data \cite{Abelev:2009wx} shows that, although they are compatible within errors, the central values for the $\pi^0$ lie systematically below the $\pi^\pm$ for low $p_T$ in the central region. The difference is much larger when nuclear modification factors of neutral pions and total charged hadrons are compared in central events, which comes basically from the baryon excess of $p/\pi$ in the intermediate momentum region $2$ GeV $<p_T<4$ GeV, where different hadron production mechanisms such as recombination come into play  \cite{Adler:2003au,Fries:2003kq}. At high enough $p_T$, say above 4-5 GeV, hadron production comes mainly from fragmentation mechanisms and the neutral- charged hadron yields tend to be similar. The experimental difficulties of accessing the low momentum region are evident and hence, the lowest point to which the yields are compared is $p_T=0.5$ GeV in the ALICE analysis \cite{ConesaBalbastre:2011zi}. At those lower momenta, the neutral-charged yields are compatible within errors, although the central $\pi^0$ value at $p_T=0.5$ GeV is slightly above the charged one. Overall, the above phenomenological data indicate compatibility with isospin symmetry within errors for the observed pion spectrum.

In this experimental context, it makes sense to explore possible differences in the charged-neutral pion masses, or more generally in their dispersion relation, which can include  momentum dependent corrections coming from thermal effects, as we will see. At the very least, this analysis should serve to confirm the very small charged-neutral deviations observed in particle distributions and would certainly be more useful to explore the low momentum region, where soft thermal pions are dominant, so that more precise experimental points at low $p_T$, as expected from ALICE data, would be welcome.

Moreover, the possible modifications in the imaginary part would give rise to differences in the thermal width between charged and neutral pions. These differences could in principle be observable at least in two phenomenological contexts. One could be  differences  in thermalization times and mean free path and hence in kinetic freeze-out temperatures for the charged and neutral pion components, estimating kinetic freeze-out as the temperature for which the mean free path becomes of the order of the system size, or equivalently for the mean collision time \cite{FernandezFraile:2009kt,Kataja:1990tp}. The other one is in transport coefficients, for which the inverse thermal width of the internal lines enters in the integrals of the relevant loop diagrams \cite{FernandezFraile:2009mi}. If there are significative differences between charged and neutral thermal widths, there could be sizable corrections e.g. to the electrical conductivity, related to the photon spectrum \cite{FernandezFraile:2005ka} or to the shear and bulk viscosities needed to explain correctly observables such as the elliptic flow or the trace anomaly \cite{FernandezFraile:2008vu,FernandezFraile:2009mi,Dobado:2008vt}.

We also recall that electromagnetic differences in meson masses at zero temperature have been measured in the lattice with increasing accuracy up to very recently \cite{deDivitiis:2013xla}. Finite temperature isospin-breaking analysis in the light quark sector are not available as far as we know, but presumably they could be affordable in the near future given the  level of precision reached  in the evaluation of finite-temperature screening properties of meson correlators \cite{Cheng:2010fe}.

Besides the possible phenomenological implications, there are other, more theoretical, aspects of our analysis, mostly in connection with chiral symmetry restoration and resonance saturation. At $T=0$, in the soft pion limit, i.e. vanishing external pion four-momentum $p_\pi$ (consistent only in the chiral limit of vanishing pion masses), and to leading order in $e^2$, the following sum rule connects the EM pion mass difference with the vector-axial spectral function difference \cite{Das:1967it}:

\begin{equation}
\lim_{p_\pi\rightarrow 0} \Delta M_\pi^2=\lim_{p_\pi\rightarrow 0} \left(\mpichsq-\mpineusq\right)=-\frac{3e^2}{16\pi^2F_\pi^2}\int_0^\infty ds \ln s \left[\rho_V(s)-\rho_A(s)\right]
\label{dasrule}
\end{equation}

A natural question in this context is therefore the possible connection  to chiral symmetry restoration at finite temperature. Since  vector and axial channels (saturated by the $\rho$ and $a_1$ resonances respectively) are meant to degenerate at the transition, the pion mass difference could then behave as an order parameter. However, as pointed out first in \cite{Kapusta:1984gj}, at finite temperature, the charged pion mass always receives a contribution $\Delta M^2(T)\sim e^2 T^2/4$, similarly to Debye screening for the longitudinal photon field, which actually would make the pion mass difference grow instead. That contribution alone would be comparable to the $T=0$ value near $T_c$. However, when the sum rule (\ref{dasrule}) is corrected at $T\neq 0$ one has to take into account also the modifications of the spectral functions $\rho_{V,A}\rightarrow \rho_{V,A}(T)$, which in the chiral limit and to leading $T^2$ order are given simply by a multiplicative $T$-dependent renormalization that mixes the vector and axial spectral functions, predicting that they become degenerate at $T\simeq \sqrt{3}F_\pi$ \cite{Dey:1990ba}. That term gives rise to a decreasing correction to $\Delta M^2(T)$ which added to the Debye-like one gives a net very soft decreasing behaviour for the pion mass difference, in agreement with the ChPT calculation in the chiral limit \cite{Manuel:1998sy}.

All these aspects already studied in the chiral limit are meant to change considerably when nonzero physical pion masses are considered. First of all, the soft pion limit will not be applicable because it only makes sense in the chiral limit. Second, for the relevant temperatures involved near chiral restoration and in heavy-ion collisions, $T$ and $M_\pi$ effects are comparable, so that new mass-dependent and momentum-dependent terms are expected, which could change the previous chiral restoring and not-restoring balance. One of our purposes in this work will be precisely  to analyze those aspects related to the connection of the self-energy electromagnetic difference with the vector and axial spectral functions when the pion mass is taken at its physical value.

Moreover, since the previous sum rule arguments and their finite-$T$ extensions are not directly  applicable out of the chiral limit, we will find useful also to  appeal  to models based on resonance exchange, for which the pion mass difference has been calculated at $T=0$  \cite{Donoghue:1996zn},  in order to identify the leading and subleading contributions for physical masses in the resonance saturation limit. Also within this framework we will establish the validity limit of our pure-ChPT calculation.

 Taking all these considerations into account, the structure of this work is the following: In section \ref{sec:chpt} we will carry out the ChPT analysis of the self-energy, the real part contribution being discussed in section \ref{sec:real} and the imaginary one in section \ref{sec:im}. In both cases we will discuss several aspects such as the differences with the chiral limit, gauge invariance, the momentum dependence and possible phenomenological consequences. Section \ref{sec:reso} will be devoted to the discussion of the extension of the sum rule connecting the pion electromagnetic self-energy difference with the vector-axial vector spectral function difference. We will review the main aspects of previous derivations, both at $T=0$ and $T\neq 0$ in the chiral limit and analyze the main differences arising for physical pion masses and the formal implications regarding chiral symmetry restoring and not-restoring terms. In section \ref{sec:expres} we will consider the pion self-energy calculation in a resonance saturation framework including the $\rho$ and $a_1$ resonances explicitly and will examine the size of the different corrections within the context of the present work. In Appendices \ref{app:genprop} and \ref{app:loop} we will clarify several properties, definitions and conventions used throughout this work regarding spectral functions and loop integrals.

\section{ChPT analysis for physical pion masses}
\label{sec:chpt}

The effective chiral lagrangian up to fourth order in $p$ (a meson mass, momentum, temperature or derivative) including EM interactions proportional to $e^2$ is given schematically by ${\cal L}_{eff}={\cal L}_{p^2+e^2}+{\cal L}_{p^4+e^2p^2+e^4}$.  The second order lagrangian corresponds to the familiar non-linear sigma model plus the addition of the gauge coupling of mesons to the photon field via the covariant derivative, and an additional term proportional to a low-energy constant $C$ compatible with the $e\neq 0$ symmetries of the QCD lagrangian \cite{Urech:1994hd,Ecker:1988te,Knecht:1997jw,Meissner:1997fa}.

\begin{equation}
{\cal L}_{p^2+e^2}=\frac{F^2}{4} \tr\left[D_\mu U^\dagger D^\mu U+2B_0{\cal M}\left(U+U^\dagger\right)\right]+C\tr\left[QUQU^\dagger\right].
\label{L2}
\end{equation}

Since we are dealing only with pions, the Goldstone Boson field matrix takes the form $U(x)=\exp [i\Phi/F]\in SU(2)$, being

\begin{equation}
\Phi= \left(\begin{array}{ll}
 \pi^0 &  \sqrt{2}\pi^+ \\
 \sqrt{2}\pi^- & - \pi^0
    \end{array}\right)
\end{equation}
the pion field matrix.

The covariant derivative is $D_\mu=\partial_\mu+iA_\mu[Q,\cdot]$ with $A_\mu$ the EM
field and $Q=(e/3)\diag(2,-1)$  and ${\cal M}=\hat m\I_2$ are respectively the quark charge and mass matrices, where we will take the QCD isospin limit $m_u=m_d=\hat m$, since as explained in the introduction, we are interested in the dominant EM isospin breaking effect in the pion masses. Thus, from the lagrangian (\ref{L2}) we read off the tree level neutral and charged pion masses, which we denote by a hat:

\begin{eqnarray}
\mpichsqzero&=&2\hat m B_0 + 2 C \frac{e^2}{F^2},\nonumber\\
\mpineusqzero&=& 2\hat m B_0. \label{treemassessu2}
\end{eqnarray}

The above squared tree level pion masses are then, consistently, $\Od(p^2)$ quantities independent of temperature, which are related to the physical pion masses formally as $M_\pi^2=\hat M_{\pi}^2+\Od(p^4)$. We will be interested here in the calculation of those $p^4$ corrections, since they include the leading temperature dependence coming from pion loops. Similarly, the pion decay constant $F_\pi^2=F^2+\Od(p^4)$ and the quark condensate $\condtwo=2B_0F^2[1+\Od(p^2)]$.

Physical predictions are rendered UV finite by renormalization of the low-energy constants (LEC) multiplying the different terms of the lagrangian. Thus, the fourth order lagrangian consists of all possible terms compatible with the QCD symmetries to that order, including the EM ones, and can be found for SU(2)-ChPT, for instance, in  \cite{Knecht:1997jw}. It introduces a set of EM and non-EM LEC which appear in the calculation of the masses by instance of ${\cal L}_{p^4+e^2p^2+e^4}$, when renormalizing the $T=0$ divergences coming from the loops.

At finite temperature $T\neq 0$, we will work in the Imaginary Time (IT) formalism \cite{galekapustabook,lebellac} in which the correlators corresponding to propagators are obtained by replacing in the action $t\rightarrow -i \tau$, $i\int{d^4x}\rightarrow \int_T d^4x\equiv\int_0^\beta d\tau \int d^3\vec{x}$. The vertices remain the same as at $T=0$ and the Feynman rules are modified according to the replacements indicated in (\ref{itrep}). Once the internal loop sums over Matsubara frequencies $\omega_n$ are performed, the result for a given correlator can be analytically continued to external frequencies $\omega+i\epsilon$ to obtain the retarded propagator, which contains the information about the dispersion relation. The details and definitions of the various propagators and spectral functions are given in Appendix \ref{app:genprop} while in Appendix \ref{app:loop} we collect the results for the typical thermal loop integrals that we will need throughout this work.

The dispersion relation is set up by the poles of the retarded propagator at $\omega=\pm\omega_p-i\gamma_p$ with $\gamma_p$ the damping rate in the thermal bath. It is obtained from the self-energy $\Sigma$, which in imaginary time is defined in (\ref{fullpropit}). As we will work perturbatively within ChPT, $\Sigma=\Od(p^4)$ so that the dispersion relation is perturbed around the vacuum value, i.e, $\omega_p^2=E_p^2+\re\Sigma(E_p;\modp;T)$ and $\gamma_p=-\im\Sigma(E_p+i\epsilon;\modp;T)/(2E_p)$ where $E_p^2=\modp^2+\hat M^2$, $\hat M$ is the tree level mass and the $T$-dependent and $\vec{p}$ dependent self-energy has been analytically continued from the IT one as $i\omega_n\rightarrow \omega+i\epsilon$ to obtain the retarded propagator $G_R(\omega,\modp)$ (see Appendix \ref{app:genprop} for more details). Recall that at $T\neq0$, $\Sigma(\omega,\modp)$ depends   separately on $\omega$ and $\modp$ as a result of the Lorentz Symmetry breaking due to the choice of the reference system associated with the thermal bath.

\begin{figure}
\centering
\includegraphics[scale=0.15]{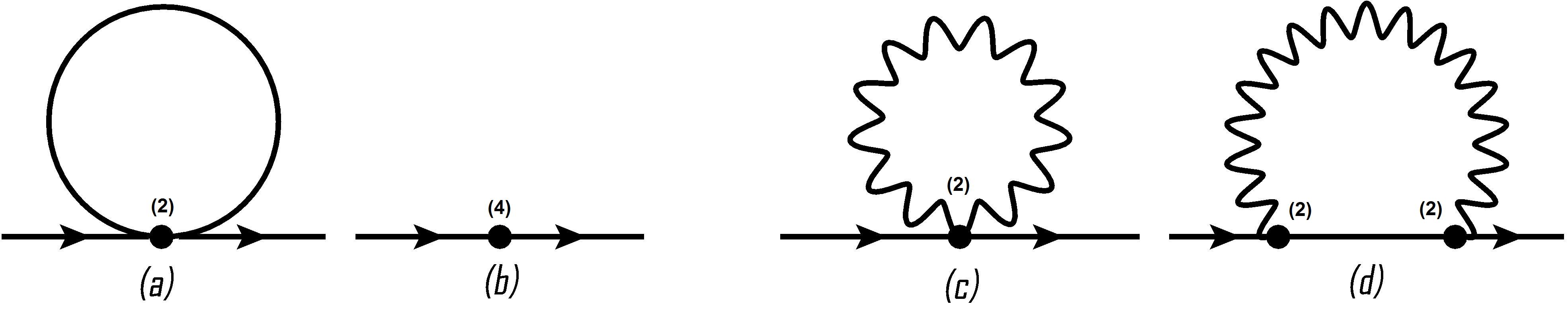}
\caption{1-PI diagrams contributing to the self-energy of a charged pion in SU(2)-ChPT to leading order. Diagrams for neutral pions are the same removing those in which photons are present.}
\label{fig:diag}
\end{figure}

If EM isospin breaking is considered, the dispersion relation is different for charged and neutral pions already at tree level, as indicated in (\ref{treemassessu2}). To one loop, the diagrams contributing to the charged pion self-energy in ChPT are showed in Fig.\ref{fig:diag}. To the neutral pion self-energy, only diagrams of type (a) and (b) contribute. The numbers between brackets in those diagrams denote the momentum order of the lagrangian that gives the corresponding vertex. Diagrams (c) and (d) involve virtual photons. It is important to note that apart from the charged-neutral differences in the self-energy coming from diagrams (c) and (d), there are others contributing at the same chiral order from diagrams of tadpole type (a). Thus, on the one hand, a four-pion vertex coming from the $F^2$ term in (\ref{L2}) gives rise to a contribution of the type $G(\hat M_{\pi^\pm})-G(\hat M_{\pi^0})$ to the self-energy charged-neutral difference, with $G$ the tadpole function defined in (\ref{Gfun}). On the other hand, a four-pion vertex coming from the $C$ term in  (\ref{L2}) only contributes to the charged pion self-energy proportionally to $C G(\hat M_{\pi^\pm})$.

We will discuss separately the real and imaginary contributions to the pion self-energy within our ChPT approach. We will refer to the Appendices for details of the calculation. All the $T=0$ contributions will be regularized in the dimensional regularization (DR) scheme throughout this work.

\subsection{Real part of the dispersion relation. Pion mass difference and momentum dependence}
\label{sec:real}

As discussed above, the real part of the self-energy shifts the real part of the pion pole, introducing $T$ and momentum dependence,  perturbatively within ChPT through  $\re\Sigma(E_p,\modp;T)$ with $E_p^2=\modp^2+\hat M^2$ with $\hat M^2$ the corresponding tree level pion mass.  As customary, we will define the pion masses in the static limit $\vec{p}=\vec{0}$.

The photon-tadpole diagram in Fig.\ref{fig:diag}(c) defines the thermal Debye screening mass for longitudinal modes \cite{galekapustabook} and vanishes at $T=0$. It is UV finite as it should for a pure thermal contribution. The UV divergences coming from the tadpole diagrams (a) and the photon-exchange diagram (d) are the same as at $T=0$ and are absorbed by the tree level diagrams (b), which include a particular combination of the fourth order LEC. The $T=0$ result for the neutral and charged pion masses taking into account all these diagrams is given in \cite{Knecht:1997jw}.

\begin{figure}
\centering
{\includegraphics[scale=0.95]{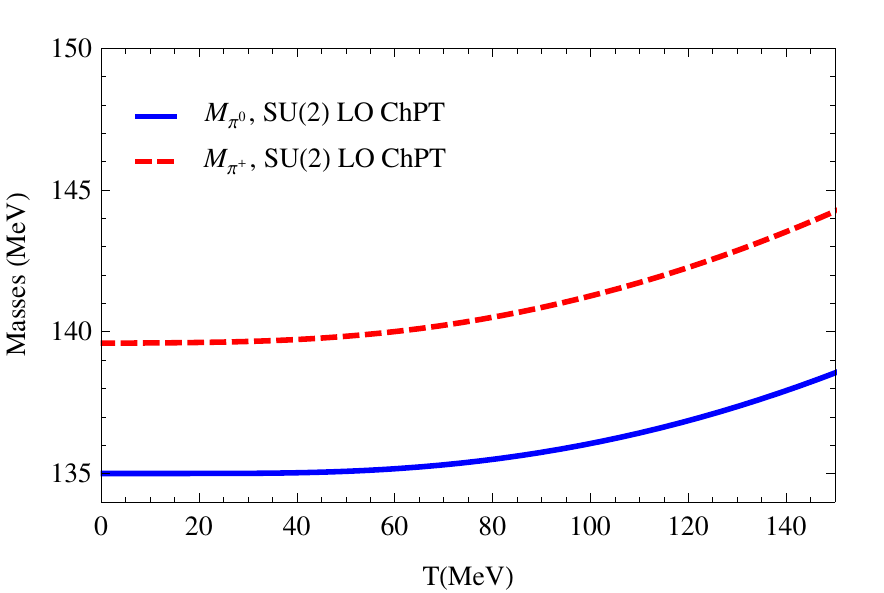}\includegraphics[scale=1.03]{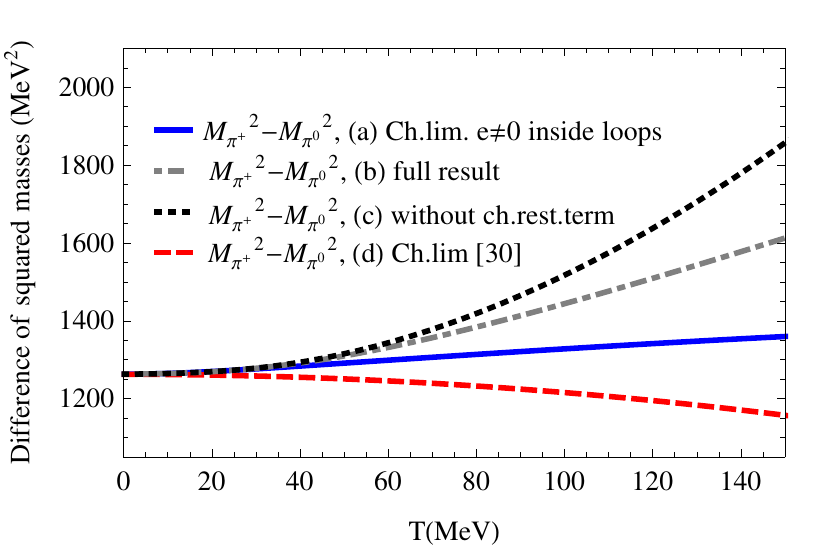}}
\caption{Left: Charged (red,dashed line) and neutral (blue,solid line) pion masses in the static limit to leading order in SU(2)-ChPT for non zero tree level pion masses. Right: Different results for the charged-neutral pion mass difference: (a, solid line) corresponds to our calculation in the chiral limit keeping  $e\neq 0$ for the tree level charged pion mass inside the loops; (b, dot-dashed line) corresponds to the full ChPT calculation with $m\neq 0$ and $e\neq 0$ also inside the loops; (c, dotted line) is the full result subtracting the chiral restoring term as explained in the text and (d,   dashed line) is the chiral limit result neglecting $\Od(e^4)$ as given in \cite{Manuel:1998sy}. }
\label{fig:masassu2chptstatic}
\end{figure}

For the neutral pion mass, the above mentioned tadpole diagrams give to this order:

\begin{equation}
\label{neutral}
M^2_{\pi^0}(T)=\hat M^2_{\pi^0}(T=0)\left[1+\frac{1}{F^2}\left(g_1(\hat M_{\pi^\pm},T)-\frac{1}{2}g_1(\hat M_{\pi^0},T)\right)\right],
\end{equation}
with the thermal $g_1$ function defined in (\ref{g1}). The above thermal neutral pion mass is still increasing with temperature, as showed in Fig.\ref{fig:masassu2chptstatic}.

As for the charged pion self-energy, for the photon-tadpole contribution in Fig.\ref{fig:diag}(c) and the photon-exchange diagram (d) we will work in the Feynman gauge $\alpha=1$ (see our notation for thermal propagators in Appendix \ref{app:genprop}) as in the $T=0$ analysis \cite{Knecht:1997jw,Donoghue:1996zn} and previous $T\neq 0$ ones \cite{Kapusta:1984gj,Manuel:1998sy,Ladisa:1999wx}. In that gauge we have for diagram (c):

\begin{equation}
\Sigma_{\gamma Tad}(T)=4e^2T\sum_n \int\frac{d^3\,q}{(2\pi)^3} \frac{1}{\omega_n^2+\vert\vec{q}\vert^2}=4e^2g_1(0;T)=\frac{e^2 T^2}{3}
\label{phtad}
\end{equation}
where $\omega_n=2\pi nT$ is the internal Matsubara frequency and we have used (\ref{g1}) and (\ref{g1asymhighT}). As commented above, this is the typical $e^2T^2$ screening or Debye mass behaviour appearing for longitudinal photon fields in the thermal bath \cite{Kraemmer:1994az,galekapustabook} which holds also for gluons with  prefactor corrections.  Note also that this is a growing term with $T$, behaving then against the naive arguments of chiral restoration mentioned in section \ref{sec:intro}.

The photon exchange term corresponding to diagram (d) in Fig.\ref{fig:diag} is given in the Feynman gauge as:

\begin{eqnarray}
\label{sigma1pe}
\Sigma_{\gamma Ex}(i\omega_m,\modp;T)=e^2T\sum_n \int\frac{d^3\,q}{(2\pi)^3} \frac{(2p-q)^2}{q^2\left[(p-q)^2-\hat M_{\pi^\pm}^2\right]}
\end{eqnarray}
where $p\equiv(i\omega_m,\modp)$ and $q\equiv(i\omega_n,\vec{q})$ are the external and loop IT momenta respectively, with $\omega_k=2\pi k T $.

Writing in (\ref{sigma1pe}), $2p\cdot q=-(p-q)^2+q^2+p^2$, we have in IT:

\begin{equation}
\Sigma_{\gamma Ex}(i\omega_m,\modp;T)=e^2\left\{G(\hat M_{\pi^\pm};T)-2G(0;T)+2\left[\hat M_{\pi^\pm}^2-\omega_m^2-\modp^2\right] J_T(0,\hat M_{\pi^\pm};i\omega_m,\modp) \right\}
\label{sigmapeJ}
\end{equation}
where the $G$ and $J_T$ functions are defined in (\ref{Gfun}) and (\ref{JTm1m2}) respectively. Therefore, performing the analytical continuation $i\omega_m\rightarrow p_0+i\epsilon$ and for on-shell pions $p^2=\hat\mpichsq$ (perturbative self-energy), this contribution   can be cast as:

\begin{equation}
\Sigma_{\gamma Ex}(\omega+i\epsilon,\omega^2=E_p^2;T)=e^2\left[G(\hat M_{\pi^\pm};T)-2G(0;T)+4\hat M_{\pi^\pm}^2 J_T(0,\hat M_{\pi^\pm};\omega+i\epsilon,\omega^2=E_p^2)\right]
\label{sigma1pe2}
\end{equation}

Note that in the chiral limit ($\hat m=0$) and neglecting $\Od(e^4)$ we get $\Sigma_{\gamma Tad}+\Sigma_{\gamma Ex}=\frac{e^2T^2}{4}$ which is nothing but the scalar thermal mass squared obtained to one loop in Scalar QED (SQED) \cite{Kraemmer:1994az}. Note also that, according to our analysis in Appendix \ref{app:loop}, the above $J_T$ function develops an imaginary part, which we will analyze in section \ref{sec:im}.

At this point, let us discuss the gauge invariance of the previous result in a covariant gauge. The gauge parameter dependence is in  the photon propagator and then it only affects diagrams (c) and (d) in Fig.\ref{fig:diag}. If we add the contribution proportional to $(\alpha-1)$ of the gauge boson propagator (\ref{photpropcov}), we obtain the following additional contributions to those diagrams

\begin{eqnarray}
\delta\Sigma_{\gamma Tad}(T)&=&-e^2(\alpha-1)\frac{T^2}{12}\nonumber\\
\delta\Sigma_{\gamma Ex}(i\omega_m,\modp;T)&=&e^2(\alpha-1) T\sum_n \int\frac{d^3\,q}{(2\pi)^3} \frac{\left[(2p-q)\cdot q\right]^2}{(q^2)^2\left[(p-q)^2-\hat M_{\pi^\pm}^2\right]}
\end{eqnarray}

Now, let us concentrate on the on-shell point $p^2=\mpichsq$ (which will hold  after analytical continuation) so that with similar manipulations as before, we can write the numerator of $\delta\Sigma_{\gamma Ex}$ as $(2p\cdot q)\left[-(p-q)^2+\mpichsq-q^2\right]+(q^2)^2=\left[-(p-q)^2+\mpichsq\right]\left[2p\cdot q-q^2\right]$ so that we get $\delta\Sigma_{\gamma Ex}(\omega^2=E_p^2)=e^2(\alpha-1)\frac{T^2}{12}=-\delta\Sigma_{\gamma Tad}(T)$ since the sum and integration of $p\cdot q/(q^2)^2$ vanishes. Therefore, within our perturbative ChPT scheme, the dispersion relation is independent of the gauge parameter in covariant gauges. Note that it is crucial that we remain within the strict regime of perturbation theory to prove this result, since, consistently with that approach, we have taken the self-energy at the on-shell point.

Therefore, we get, after collecting  all the pieces, for the real part of the self-energy difference at finite temperature:
\begin{eqnarray}
\Delta\Sigma (\modp;T)&=&\Delta\Sigma (T=0)+\frac{\hat M^2_{\pi^0}}{F^2}\left[g_1(\hat M_{\pi^0},T)-g_1(\hat M_{\pi^\pm},T)\right]+(1-4Z)e^2g_1(\hat M_{\pi^\pm},T)\nonumber\\&+&\frac{e^2T^2}{6}+4\mpichsq\re J_T(0,\hat M_{\pi^\pm};\modp) + \Od(p^6),
\label{selfchpt}
\end{eqnarray}
where the explicit expression for  $\re J_T(0,\hat M_{\pi^\pm};\modp)$ is given in (\ref{rejtphex}) and $Z=C/F^4$. We recover the $T=0$ result of \cite{Knecht:1997jw} taking into account (\ref{Gfunsep}) and (\ref{rejtphexT0}). On the other hand, in the chiral limit $\hat m=0$ and neglecting $\Od(e^4)$, we reproduce the result of \cite{Manuel:1998sy} for the EM mass difference, namely $M^2_{\pi^\pm}-M^2_{\pi^0}=(2Ce^2/F^2)\left(1-\frac{T^2}{6F^2}\right)+\frac{1}{4}e^2T^2.$ Note the combination of a first decreasing term, coming from vector-axial mixing towards chiral restoration (as we will discuss in section \ref{sec:reso}) plus the increasing thermal scalar mass term. The net result in the chiral limit is a slowly decreasing function as showed in Fig.\ref{fig:masassu2chptstatic}.

In our present work, we have additional mass and momentum dependence terms, which should play a relevant role for the physically realistic temperature regime, where the approach $T\gg M_\pi$ is not justified. In particular, if we define the mass in the static limit $\vec{p}=\vec{0}$, using  (\ref{rejtphexp0}) in (\ref{selfchpt}) we get:

\begin{eqnarray}
\Delta M_\pi^2 (T)&\equiv& \Delta\Sigma (\modp=0;T)=\Delta M_\pi^2 (0)+\frac{\hat M^2_{\pi^0}}{F^2}\left[g_1(\hat M_{\pi^0},T)-g_1(\hat M_{\pi^\pm},T)\right]+(1-4Z)e^2g_1(\hat M_{\pi^\pm},T)\nonumber\\&+&\frac{e^2T^2}{6}
+4\mpichsq g_2(M_{\pi^\pm};T)+\Od(p^6)\nonumber\\
&=&\Delta M_\pi^2 (0) +e^2\left[2(2+Z) \hat M_\pi^2 g_2(\hat M_\pi;T)+(1-4Z)g_1(\hat M_\pi;T) + \frac{T^2}{6}\right] + \Od(e^4)  +\Od(p^6)
\label{selfchptp0}
\end{eqnarray}
with $g_2$ defined in (\ref{g2}). Note that, as it is written in the last line in (\ref{selfchptp0}), it is clear that all the terms give  contributions increasing with $T$ except for  the term $-4Zg_1$ which, as we will show in section \ref{sec:reso} carries out the chiral-restoring $V-A$  mixing. Note also that apart from the $g_1(M,T)$ terms which become $T^2/12$ in the chiral limit, there is also a $g_2$ term which was absent in the chiral limit and receives a contribution from the photon exchange diagram and another one from the tadpole difference of $g_1$ functions. Taking into account the typical asymptotic behaviours for these functions described in Appendix \ref{app:loop}, this new $g_2$ term is comparable to the others for the range of temperatures considered here, i.e., relevant for a Heavy Ion environment and actually, as we will just see, the net result for the pion mass difference is now an increasing function of $T$. Recall that both $g_1$ and $g_2$ are increasing functions of $T$.

The results for the charged and neutral masses separately and for their difference are displayed in Fig.\ref{fig:masassu2chptstatic}. We have limited the temperature range to $T=150$ MeV, the typical validity range for finite-temperature ChPT calculations, i.e. $T\lsim M_\pi$.  For the numerical evaluation of our results we will take the same values for the low-energy constants as in \cite{Nicola:2010xt}. We have used physical masses for the pions instead of the tree level masses for the numerical results since the difference is encoded in higher order corrections. In the thermal range considered, and despite the different sign of the various terms, the increasing terms  turn out to dominate the pion mass difference, which is approximately $24\%$ bigger at $T=150$ MeV than the zero temperature value and altogether the variation is quite soft with temperature. Also, when $T$ grows much above the applicability range of these ChPT calculations the mass difference starts to decrease. But this should be expected since expansions in $M_{\pi^\pm}/T\rightarrow 0$ should coincide with the $T$-decreasing chiral limit behaviour commented above. It is important to remark though that for low and moderate temperatures our result with  physical masses differs  qualitative and quantitatively from the chiral limit one. Finally, we have shown also in the right panel of Fig.\ref{fig:masassu2chptstatic} the results in a modified chiral limit where we set $\hat m=0$
but consider EM effects to be still turned on, i.e. $e\neq 0$, even inside the loops. In addition, in order to calibrate the importance of chiral symmetry restoration in the obtained behaviour,   we have also plotted in Fig.\ref{fig:masassu2chptstatic} the result for the EM (static) mass difference without including the chiral restoring term $-4Zg_1(M_\pi,T)$ in (\ref{selfchptp0}). The effect  would be much larger then, giving rise to about a 6.8 MeV mass difference around $T=150$ MeV, i.e. about 1.5 times its $T=0$ value.

One of the conclusions of this work is then that the  scalar mass inherent to the thermal bath plus the massive pion effects overshadow the restoring terms coming from axial-vector degeneration leaving no trace of a chiral restoring behaviour as would have been inferred naively from (\ref{dasrule}). On the contrary, the net result is monotonically increasing. In  section \ref{sec:reso} we will present  a more detailed discussion in connection with sum rules and resonance saturation.

Let us  analyze now the momentum dependence in the real part of the dispersion relation. The pion gas formed after a relativistic heavy ion collision is in thermal equilibrium and hence momenta are weighted with the Bose-Einstein distribution function.  Thus, we can define a momentum-averaged mass and compare with the static mass defined before. This is  then a  relevant observable  when comparing with experimental pion distributions. The distribution function peaks around some three-momentum value which varies with temperature, in such a way that for a certain $T$ there are only an effective number of pions with three-momenta around this value which are thermally active. Actually, for small $T\ll M_\pi$, momenta are distributed around $p\sim \sqrt{MT}$ while in the opposite regime $T\gg M$ they do around $p\sim T$.

For any $\vec p$-dependent observable, $\mathcal{A}(\vec p,T)$, we can associate a momentum average taking into account the neat effect of the thermal bath by weighting over the number of particles present at a given temperature and dividing by the total number of pions existing in the gas, i.e.

\begin{equation}
\label{mediaobs}
\langle \mathcal{A}(T)\rangle_p=\frac{\int d^3\vec{p}\; n_B(E_p,T)\mathcal{A}(\vec{p},T)}{\int d^3p\; n_B(E_p,T)}.
\end{equation}

\begin{figure}
\centering
{\includegraphics[scale=0.7]{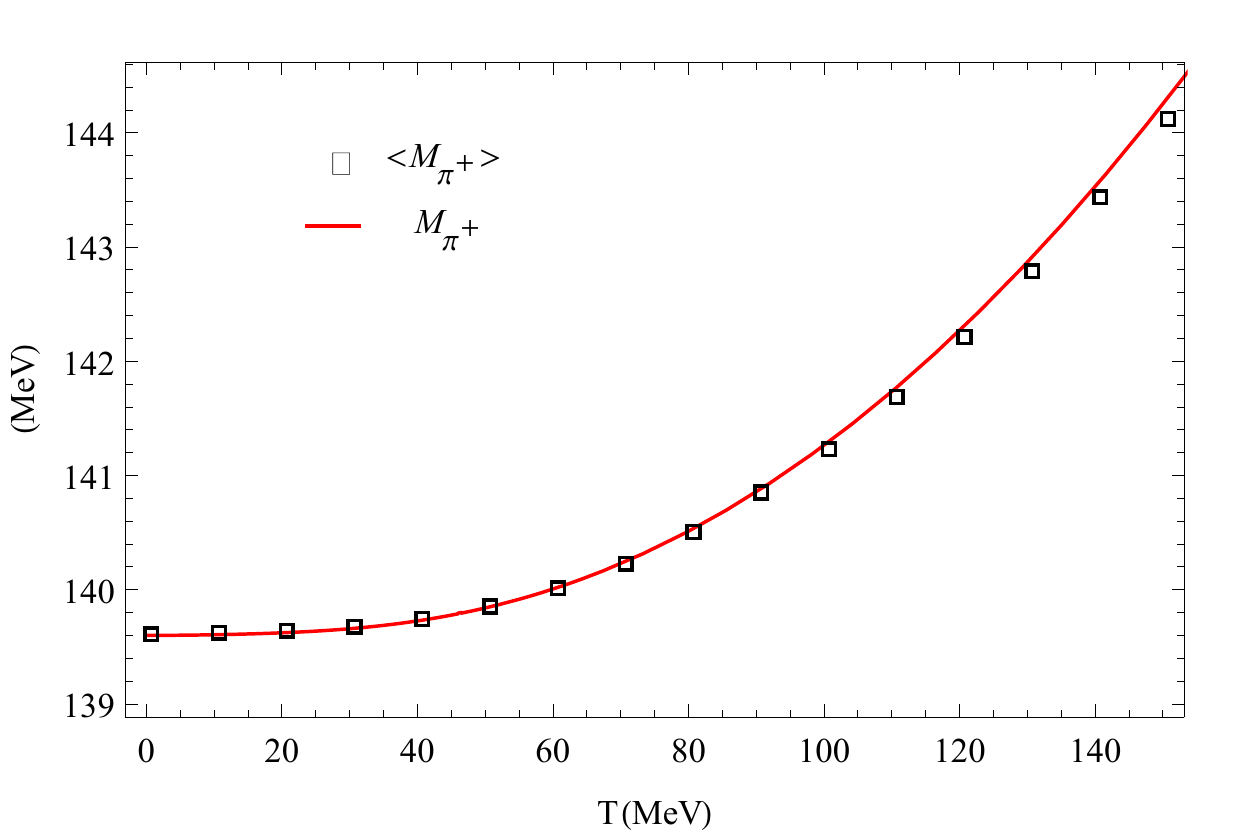}\includegraphics[scale=0.7]{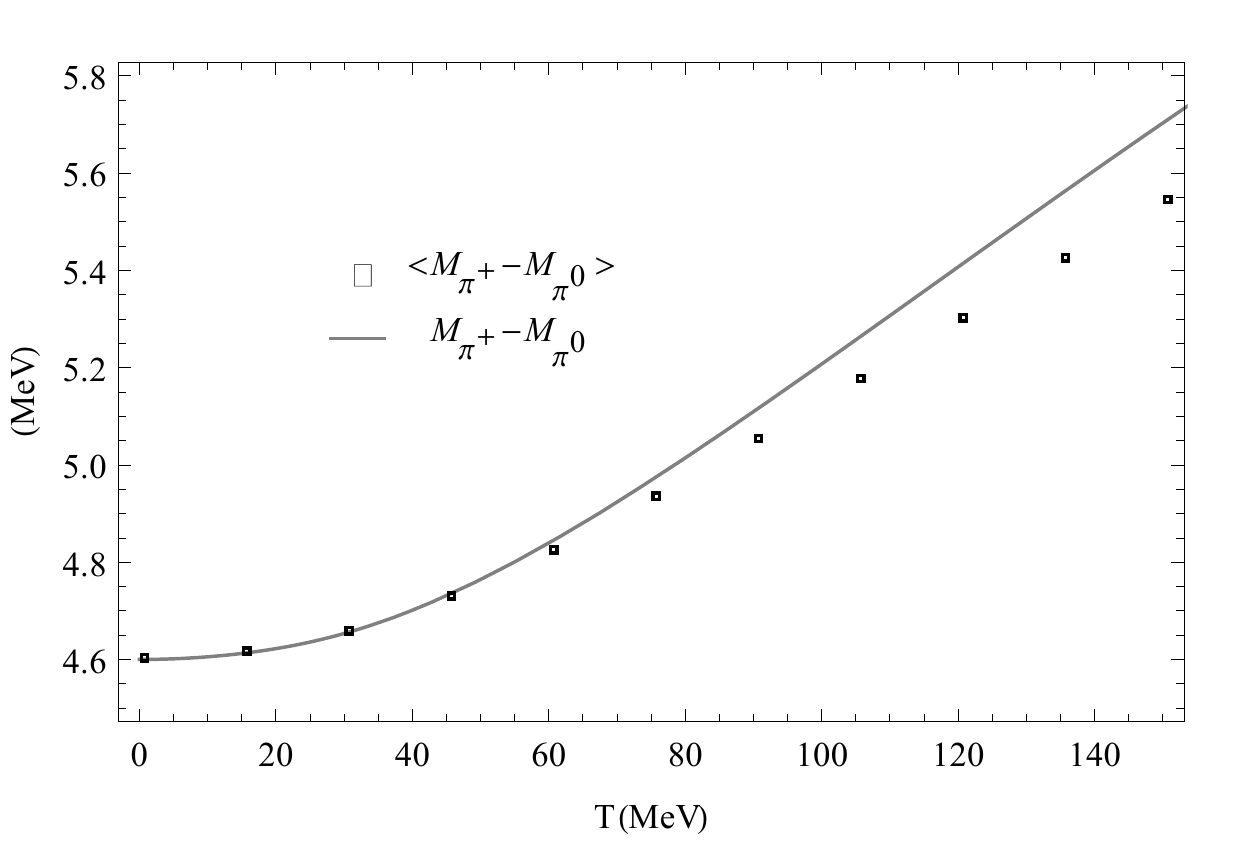}}
\caption{Leading order ChPT result for the static charged pion mass (left) and the charged-neutral pion mass difference (right) versus the mean value of those same observables over external momenta in the thermal bath.}
\label{fig:masasmedias}
\end{figure}

In Fig.\ref{fig:masasmedias} we show the results for the averaged charged pion mass (left panel) and for the charged-neutral difference (right panel)  compared with the results in the static limit. Since eq.(\ref{neutral}) does not depend on $\vec p$, neutral pions satisfy $\langle M_{\pi^0}\rangle=M_{\pi^0}$. As we see there, both pictures show that at very low temperatures the results are almost indistinguishable and, in the case of the charged mass, almost imperceptible, along the range of temperatures at which ChPT can be still predictive. The departure from the static limit is more perceptible in the mass difference since we are subtracting the main vacuum contribution to the neutral and charged masses. In that plot, note that even at moderate temperatures of about $T=$100 MeV, the effect of the thermal bath makes the averaged curve to grow slower than the static one and, for larger temperatures, we even obtain a qualitative decreasing, eventually approaching  the chiral limit behaviour faster than in the static case. Since we expect the momentum distribution to be peaked around $p\sim T$ as $T$ is increased, it is not surprise that the differences with the $p=0$ case become more relevant for higher temperatures. Note also that from (\ref{rejtphex}), we obtain that the $\re J_T$ term in (\ref{selfchpt}) vanishes asymptotically for $p\rightarrow\infty$ as $\Od(M_\pi^2/p^2)$, so that the importance of that $p$-dependent term becomes  gradually smaller as $T$ increases and therefore the total result gets closer to the chiral limit.

The main conclusion of this section is that  the EM mass difference when physical pion masses are considered is a softly increasing function of $T$, pretty much as in the $e=0$ case. This behaviour is even softer for the momentum averaged mass. This result is consistent with the experimental observations in the pion spectra commented in section \ref{sec:intro}. In this respect,  one can actually consider chiral symmetry restoration as being ultimately  responsible for charged-neutral differences not being observed, in view of the results showed in Fig.\ref{fig:masassu2chptstatic}.

\subsection{Imaginary part: bremsstrahlung-like IR enhanced contributions}
\label{sec:im}

To the order in ChPT that we are considering, the photon-exchange diagram (d) in Fig.\ref{fig:diag} leads to a nonzero imaginary part of the self-energy which, according with our previous discussion,  allows to define perturbatively a damping rate for the pion as $\gamma_{EM} (\modp)=-\im\Sigma(E_p+i\epsilon,\modp)/(2E_p)$. By the subscript EM we recall that this would be a pure EM correction felt only by the charged pions and therefore would introduce neutral-charged differences in the damping effects, as discussed below.

 In a covariant gauge, for which we have just showed that the on-shell one-loop $\Sigma$ is independent of the gauge parameter $\alpha$, and according to (\ref{sigma1pe2}), we would have  $\gamma_{EM} (\modp)=-\frac{2M_\pi^2}{E_p}\im J_T(0,M_\pi;\omega=E_p,\vert\vec{p}\vert)$ with $\im J_T$ the function given in (\ref{imjtphex}). Note that we get a nonzero answer despite the fact that the vacuum bremsstrahlung process of a scalar radiating a photon is forbidden. The reason is that, as discussed in Appendix \ref{app:loop}, the Landau and unitarity cuts in this case give a contribution for which, respectively, the conditions $E_p=\modq\pm E_{\vert\vec{q}-\vec{p}\vert}$, with $\modq$ and $\vert\vec{q}-\vec{p}\vert$ the photon and internal pion  momentum respectively, are fulfilled for $\modq=0$. Thus, those terms correspond to the two possible  processes $\pi\rightarrow\pi\gamma$ arising from cutting diagram (d) in Fig.\ref{fig:diag}, with thermal photons (quasiparticle states) weighted by $n(\modq)\sim T/\modq$ which enhances this contribution so that $\int q n(q)\delta (q)$ remains finite according to the prescription for the $\delta$ function arising from the retarded propagator, as we explain in detail at the end of Appendix \ref{app:loop}.

However, the previous covariant gauge calculation of the damping rate is not well defined. In particular, one readily realizes that $\im\Sigma$ thus obtained  is positive, so that the damping rate would be negative and then unphysical,  the corresponding retarded propagator  not having the correct analytic behavior described in Appendix \ref{app:genprop}. This sign problem is just a reflection of a deeper issue directly related to the gauge choice. For the imaginary part, we are putting the internal quasiparticles in the loop on their mass shell, weighted by the different thermal distributions. That means that in a covariant gauge, we are counting the additional nonphysical gauge degrees of freedom as being in thermal equilibrium and hence contributing to the damping rate. The problem of introducing the correct degrees of freedom in hot gauge theories has been actually treated extensively in the literature \cite{galekapustabook}. For instance, a strict loop calculation of the gauge field damping rate leads to a dependence on the gauge parameter $\alpha$ when working in covariant gauges, which may actually result in a wrong sign for the damping rate \cite{Wang:2004tg,Mottola:2009mi}. This problem is avoided in physical gauges such as the Coulomb gauge, where one gets  physically meaningful answers  \cite{Rebhan:2001wt}. To arrive to the same result in covariant gauges, alternative approaches have to be used    \cite{Landshoff:1992ne,Kraemmer:2003gd} which yield modifications of the naive gauge field propagator so as to ensure that only the physical gauge degrees of freedom remain thermally active. Actually, as it is well known in thermal field theory, these kind of difficulties with gauge invariance of the standard loop calculations was one of the motivations that led to the formulation of the Hard Thermal Loop (HTL) resummation scheme at high temperatures  \cite{Braaten:1989mz}. However, within the ChPT framework for physical pion masses, we are not in the regime where a HTL-based approach would be applicable since temperature, mass and momenta are all of the same order, so  we have to ensure that the correct degrees of freedom for thermal quasiparticles are included. For that purpose, we will define the charged pion damping rate in the strict Coulomb gauge, which free propagator is given in (\ref{photpropcoulomb}). It contains only longitudinal $D^{00}$ and transverse $D^{ij}$ components, the longitudinal one not propagating, since the corresponding free spectral function vanishes.   Note that the previous arguments should not affect the real part calculation performed in section \ref{sec:real} in covariant gauges and actually we have checked that the real part of the perturbative on-shell self-energy  remains the same in the Coulomb gauge. We also point out that previous calculations of the charged scalar damping rate in SQED, formally  similar to ours although within the HTL regime,  are also carried out  in the Coulomb gauge \cite{Thoma:1996ag,Abada:2005jq}. In those works, similarly to QCD, it is found that the transverse part of the HTL-resummed  damping rate is infrared divergent, while the longitudinal part remains finite.

It must be born in mind that the gauge problem mentioned above, as well as the existence of infrared singularities for the damping rate and a nonzero longitudinal contribution in SQED, are warnings that may indicate the necessity of considering higher terms also in our ChPT analysis, which is beyond the scope of this work. We  consider then our results in this section as mere estimates of the possible size of this pion damping effect and its consequences, which have the advantage that, at least to the order considered, the results are guaranteed to be infrared finite, as well as model-independent. The inclusion of those higher orders could actually amplify some of the phenomenological consequences that we will just discuss.

Guided by the previous considerations, we will calculate the charged pion damping rate in the strict Coulomb gauge, with gauge propagator given by (\ref{photpropcoulomb}). When this propagator is used in diagrams (c) and (d) of Fig.\ref{fig:diag}, we obtain respectively, in dimensional regularization:

\begin{eqnarray}
\Sigma_{\gamma Tad}^{CG}(T)&=&-e^2T\sum_n \int\frac{d^3\,q}{(2\pi)^3}\frac{\delta_{ij}P_T^{ij}(q)}{q^2}=-2e^2T\sum_n \int\frac{d^3\,q}{(2\pi)^3}\frac{1}{q^2}=\frac{e^2 T^2}{6} \label{phtadCG}\\
\Sigma_{\gamma Ex}^{CG}(i\omega_m,\modp;T)&=&e^2T\sum_n \int\frac{d^3\,q}{(2\pi)^3}\frac{(2\omega_m-\omega_n)^2}{\modq^2 \left[(p-q)^2-\hat M_{\pi^\pm}^2\right]}-4e^2\modp^2T\sum_n \int\frac{d^3\,q}{(2\pi)^3} \frac{1-\cos^2\theta}{q^2 \left[(p-q)^2-\hat M_{\pi^\pm}^2\right] }
\label{sigmapecg}
\end{eqnarray}
where $\cos\theta=\frac{\vec{p}\cdot\vec{q}}{\modp\modq}$.
Of the above terms, only the second one in the r.h.s. of (\ref{sigmapecg}) contributes to the imaginary part when $\Sigma$ is analytically continued and taken on the mass shell. This is precisely the transverse contribution, second term in (\ref{photpropcoulomb}), to the photon exchange diagram. The longitudinal part does not contribute to the photon spectral function at this order, consistently with the idea that longitudinal free photons do not propagate.

As commented above, we have explicitly checked that $\Sigma_{\gamma Tad}^{CG}(T)+\re\Sigma_{\gamma Ex}^{CG}(\omega^2=E_p^2,\modp;T)$ equals the result (\ref{selfchpt}). As for the imaginary part, which as stated is only well defined in the Coulomb gauge, after analytic continuation and following the same steps as in Appendix \ref{app:loop} when analyzing $\im J_T(0,M)$, we get:

\begin{eqnarray}
\gamma_{EM}(p)&=&\frac{e^2}{8\pi}\frac{p^2}{E_p^2}\int_0^\infty dq q n(q) \delta(q) \int_{-1}^1 dx (1-x^2)\left[\frac{1}{1-\frac{px}{E_p}}+\frac{1}{1+\frac{px}{E_p}}\right]\nonumber\\
&=&\frac{e^2 T}{4\pi} \left[1-\frac{M_\pi^2}{2pE_p}\log\left(\frac{E_p+p}{E_p-p}\right)\right]
\label{gammaem}
\end{eqnarray}
where we have denoted $p\equiv\modp$ and we have used the retarded prescription for the $\delta$ function discussed in Appendix \ref{app:loop}. Once more, the above contribution comes from processes radiating  thermal (physical) photon degrees of freedom at vanishing spatial momentum.

\begin{figure}
\centering
{\includegraphics[scale=.33]{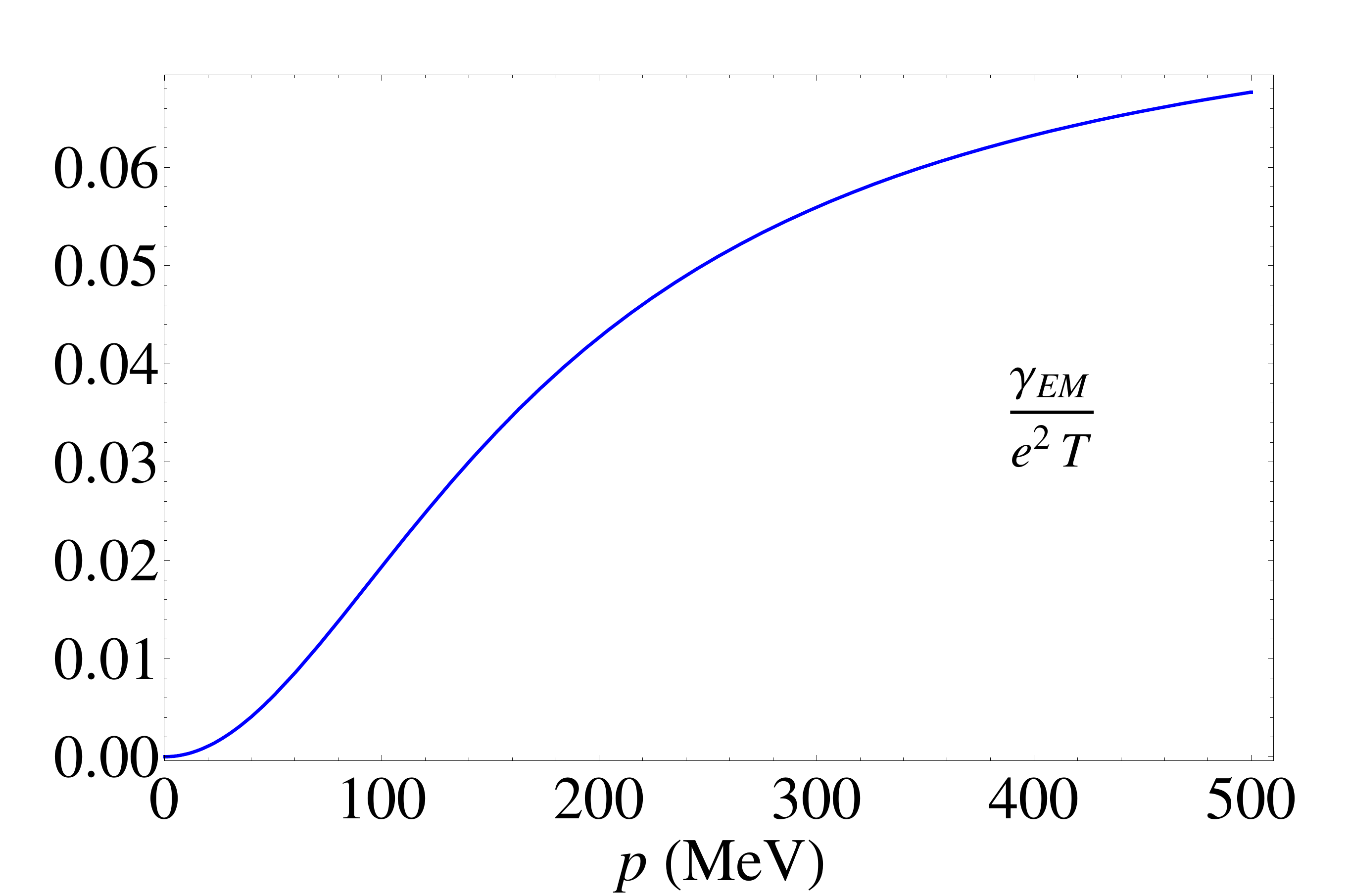}\includegraphics[scale=.3]{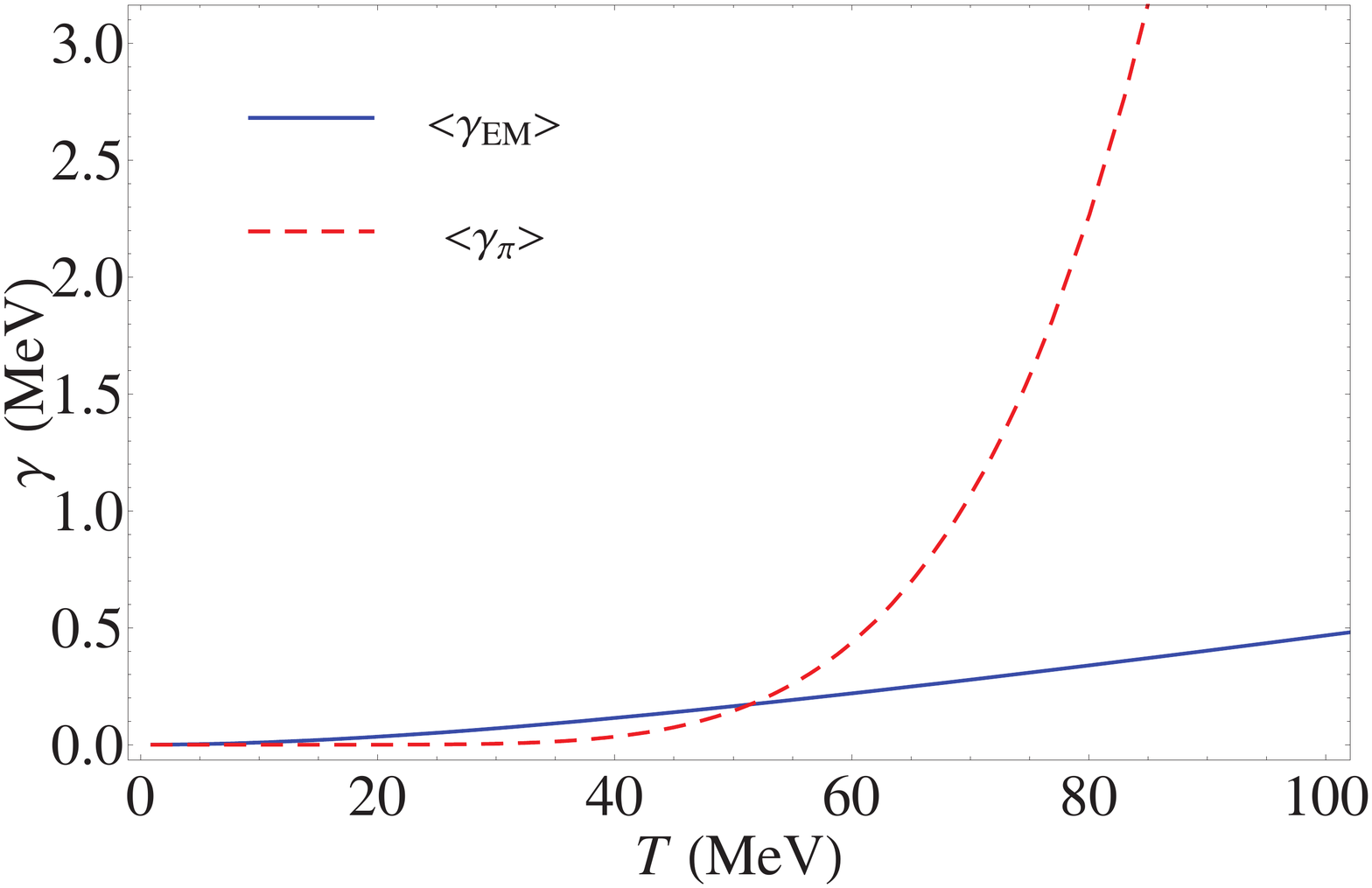}}
\caption{Left: The EM damping rate versus momentum. Right: The momentum averaged pion damping rate with $e=0$ $\langle\gamma_\pi\rangle$ and the EM contribution $\langle\gamma_{EM}\rangle$.}
\label{fig:damping}
\end{figure}

The function $\gamma_{EM} (p)$ in (\ref{gammaem}) is plotted in Fig.\ref{fig:damping} (left panel). As it can be directly checked from (\ref{gammaem}), it is linearly  proportional to $T$, it vanishes for $p\rightarrow 0^+$ for fixed pion mass $M_\pi$, and behaves asymptotically as $\gamma_{EM} (p\rightarrow\infty)\rightarrow \frac{e^2 T}{4\pi}$. This asymptotic value is also the result in the chiral limit $M\rightarrow 0^+$ or taking directly $M=0$ from the start. This $p$ dependence is indeed very similar to the one found in \cite{Thoma:1996ag} for the transverse part of the damping, although in that work $\gamma$ is logarithmically dependent on the infra-red cutoff, which we do not need to introduce at the order we are considering. In turn, we mention that we have checked that we arrive also to (\ref{gammaem}) by replacing in the general expressions in \cite{Thoma:1996ag} the free spectral function in the Coulomb gauge.

In order to analyze the possible phenomenological effects of this EM contribution to the damping rate, we have plotted in Fig.\ref{fig:damping} (right panel) the average $\langle\gamma_{EM}\rangle$ according to (\ref{mediaobs}), compared to the averaged damping rate in ChPT for $e=0$, which we denote $\gamma_\pi$  which comes from a two-loop sunset diagram \cite{Schenk:1993ru} and can be obtained also from kinetic theory arguments \cite{Goity:1989gs}. The damping $\gamma_\pi$ is the leading contribution to the  inverse mean collision time and to the inverse mean free path for pions in the isospin limit, i.e., contributes equally for  charged and neutral pions, whereas, as stated above, $\gamma_{EM}$ would contribute only to the charged ones. We also recall that $\gamma_\pi$, within the dilute regime applicable here, depends linearly on the imaginary part of the $\pi\pi$ scattering forward amplitude and hence on the total cross section from the optical theorem, which allows to get a unitarized version whose average value grows much smoothly with temperature due to the unitarity bounds on the amplitude \cite{Schenk:1993ru}. This unitarized damping is actually more realistic physically, since it describes scattering more accurately.

From the curve in Fig.\ref{fig:damping}, we observe that, even though in principle $\gamma_\pi$ is a higher order contribution with respect to $\gamma_{EM}$ in the ChPT expansion, their numerical values are comparable for low and moderate temperatures and $\gamma_\pi$ gets actually much larger as $T$ increases further. This is due on the one hand to the small numerical size of EM contributions and on the other hand to the large growing with $T$ of the nonunitarized damping discussed above, due to the strongly interacting character of pion scattering as energy increases. The second effect starts being significant from about $T\simeq$ 80 MeV, although the unitarized curve still departs  from the EM one above $T=100$ MeV.

Thus, within a Heavy-Ion environment, we expect the maximum EM effects to be operative at the end of the expansion, i.e. around thermal freeze out $T\simeq$ 100 MeV. As discussed in section \ref{sec:intro}, the thermal damping $\gamma(p)$ enters inversely in transport coefficients, inside a $p$ integral corresponding to the leading diagram for conserved current correlators \cite{FernandezFraile:2005ka,FernandezFraile:2009mi,FernandezFraile:2008vu}. It is not the purpose of this work to carry out a detailed evaluation of this effect, but in order to get a rough estimate of the size of the corrections, we can use just the thermal averaged values. In particular, in the electrical conductivity only the charged pion enters in the dominant loop \cite{FernandezFraile:2005ka} so that an estimate of the correction to that coefficient with respect to the isospin limit would be of order $1-\gamma_\pi/(\gamma_\pi+\gamma_{EM})$, which gives, for averaged values, 0.07 for $T=100$ MeV (taking the unitarized value for $\langle\gamma_\pi\rangle$) and 0.13 for $T=80$ MeV so in that region the expected correction to the electrical conductivity is around 10\%. Regarding other transport coefficients, such as the shear and bulk viscosities, since all pion species enter the loop of energy-momentum correlators \cite{FernandezFraile:2009mi}, the correction will be of order $1-[1+2\gamma_\pi/(\gamma_\pi+\gamma_{EM})]/3$ which gives  0.05 for $T=100$ MeV and 0.09 for $T=80$ MeV.

Another consequence of the EM damping effect is that the mean free time $\tau=1/\gamma$ and the mean free path $\lambda=p/(E_p \gamma)$ for charged pions are smaller than for the neutral component.  Thus, for neutral pions $\tau_0=1/\gamma_\pi$ while for charged ones $\tau_{ch}=1/(\gamma_\pi+\gamma_{EM})$. This implies for instance a reduction in the thermal or kinetic freeze-out temperature of the charged pion component with respect to the neutral one, defined as $\tau (T_{FO})\simeq 10$ fm/c, the typical plasma lifetime. This effect is much smaller: we get  around 2 MeV reduction in $T_{FO}$ between the neutral and charged components, using again the unitarized $\langle\gamma_\pi\rangle$.

\section{Sum rule, resonances and chiral restoration}
\label{sec:reso}

The study of sum rules regarding spectral functions in the vector and axial-vector channels and their in-medium or thermal bath modifications has been  the subject of thorough investigation up to very recently \cite{Kapusta:1993hq,Hohler:2012xd,Holt:2012wr}. We will be interested here in the sum rule related to the EM pion mass difference and its extension to finite pion masses and finite temperature. That sum rule was originally derived in \cite{Das:1967it} and analyzed at finite temperature in the chiral limit in \cite{Kapusta:1984gj,Manuel:1998sy}.

The traditional derivation of Das sum rule \cite{Das:1967it,Donoghue:1996zn,Manuel:1998sy} starts from the $\Od(e^2)$ correction to the pion mass given in terms of a EM current-current correlator:

\begin{eqnarray}
\Delta \Sigma (\modp;T)=\frac{e^2}{2}\int_T d^4x \langle \pi^+ (p) \vert {\cal T} J_\mu (x) J_\nu (0) \vert \pi^+ (p)\rangle_T D_0^{\mu\nu} (x)\nonumber\\
=\frac{e^2}{2}T\sum_n \int \frac{d^3\vec{q}}{(2\pi)^3}\frac{g^{\mu\nu}T_{\mu\nu}(q,p;T)}{\omega_n^2+\vert\vec{q}\vert^2}
\label{compton}
\end{eqnarray}

Here, we have allowed for a $\vec{p}$ dependence on the self-energy, from the loss of Lorentz covariance. The $\vert\pi\rangle$ states are meant to be $T=0$ free ones with dispersion relation $p^2=M_\pi^2 (e=0)$. The current $J_\mu$  is the EM current, whose $QCD$ representation is  $J_\mu=\bar q Q \gamma_\mu q$ and time ordering is along the imaginary time axis. The subscript $T$ in the matrix element indicates that the corresponding IT correlators obtained after LSZ reduction formulas are to be averaged in the thermal bath. After all the Matsubara sums are performed, the result for the self-energy defined in (\ref{compton}) is meant to be analytically continued to the external $p=(\omega+i\epsilon,\vec{p})$ with $\omega\in\IR$, so that this corresponds to the retarded self-energy, which encodes properly the spectral properties, as discussed in  Appendix \ref{app:genprop}.

Since this is just the leading order correction in $e^2$, we can take $D_0^{\mu\nu}$ as the free photon propagator, which we  consider in the Feynman gauge $\alpha=1$. $T_{\mu\nu} (q,p)$ is the Fourier transform of the pion matrix element in the first equation above and corresponds to Compton scattering. It is useful to split this amplitude into contact and non-contact  terms  $T_{\mu\nu}=T_{\mu\nu}^{(C)}+T_{\mu\nu}^{(NC)}$ so that the contact contribution corresponds to  two photons interacting in the same vertex (seagull diagram (c) in Fig.\ref{fig:diag} in our previous ChPT calculation), i.e,  $T_{\mu\nu}^{(C)}=2g_{\mu\nu}$.

\subsection{$T=0$ sum rule in the Soft Pion Limit}
\label{sec:T0sumrule}

In order to relate (\ref{compton})  with vector and axial-vector thermal averages, suitable to connect with chiral restoration, one possible approach is to take the Soft Pion Limit (SPL) $p\rightarrow 0$ for the pion states  and use Current Algebra (CA) for the current commutators involved. Note that using the SPL implies automatically to work in the chiral limit of vanishing quark masses $\hat m=0$ so that the $\pi^0$ is massless. Let us first analyze the $T=0$ case in the SPL.

In the SPL+CA approach the non-contact part of the Compton amplitude satisfies:

\begin{equation}
\lim_{p\rightarrow 0} T_{\mu\nu}^{(NC)}(q,p)=\frac{2}{F_\pi^2} \left[\Pi^V_{\mu\nu}(q)-\Pi^A_{\mu\nu}(q)\right]
\label{comptonspl}
\end{equation}
where $\Pi^{V,A}_{\mu\nu}(q)$ are respectively the Fourier transforms of the vector and axial-vector vacuum expectation values  $\langle 0\vert V_\mu^3 (x) V_\nu^3 (0)\vert 0 \rangle$ and $\langle 0\vert A_\mu^3 (x) A_\nu^3 (0) \vert 0 \rangle$ with $V_\mu^a (x)$ and $A_\mu^a (x)$ the vector and axial-vector currents. Note that here we do not make any distinction between the physical $F_\pi$ and the tree level $F$ appearing in the lowest order chiral lagrangian (\ref{L2}) since they coincide in the regime of validity of CA, equivalent to the lowest order in the chiral expansion.

 For the non-contact contribution we use the standard $T=0$ decomposition (see Appendix \ref{app:genprop}):

\begin{eqnarray}
\Pi^V_{\mu\nu}&=&\left(\frac{q_\mu q_\nu}{q^2}-g_{\mu\nu}\right)\Pi^V (q^2)\nonumber\\
\Pi^A_{\mu\nu}&=&\left(\frac{q_\mu q_\nu}{q^2}-g_{\mu\nu}\right)\Pi^A_t (q^2) + \frac{q_\mu q_\nu}{q^2} \Pi^A_l (q^2)
\end{eqnarray}

Note that for the axial-vector case, we have added a four-dimensional longitudinal piece, which arises from the partial conservation of axial current (PCAC) in QCD. We use  $T,L$ to denote three-dimensional transverse and longitudinal contributions (both four-dimensionally transverse) and $t,l$ to denote four-dimensional transverse and longitudinal ones.

On the other hand, as customary, we can write for the correlators $\Pi^V$ and $\Pi^A_t$ their  spectral function representation at $T=0$:

\begin{eqnarray}
\Pi^{V}(q^2)&=& q^2 \int_0^\infty ds \frac{\hat \rho^V (s)}{q^2-s}\nonumber\\
\Pi_t^{A}(q^2)&=& q^2 \int_0^\infty ds \frac{\hat \rho^A (s)}{q^2-s}
\end{eqnarray}
since at $T=0$ they only depend on $q^2$ and there are no cuts for $s<0$  (see Appendix \ref{app:genprop}).

To leading order in the low-energy expansion, or equivalently using CA, in the chiral limit and for $T=0$ one has $\Pi^A_l (q)=F_\pi^2 q^2 G_\pi (q)$ with $G_\pi$ the leading order pion propagator, since the axial-vector current in the low-energy representation, from (\ref{L2}), is just $A_\mu^a=F_\pi\partial_\mu \pi^a + \dots$, the dots denoting  higher terms in the chiral expansion (labeled formally by the $1/F$ in the lagrangian). This is consistent also with the PCAC theorem, valid within CA, $\langle 0 \vert  \partial^\mu A_\mu^a \vert \pi^b\rangle=\delta^{ab} F_\pi M_\pi^2$.

Thus, for $T=0$ and in the chiral limit one has:

\begin{equation}
\left.\Delta M_\pi^2\right\vert_{SPL, T=0}=\frac{3e^2}{F_\pi^2} i \int \frac{d^4 q}{(2\pi)^4} \frac{q^2}{q^2+i\epsilon}\left[\frac{F_\pi^2}{q^2+i\epsilon} -\int_0^\infty ds \frac{\rho^V(s)-\rho^A(s)}{q^2-s+i\epsilon}\right]
\label{sumruleT0SPLprev}
\end{equation}
where the momentum integral is in Minkowski space-time. The first term inside the brackets in the above expression comes from the sum of the contact term $T_{\mu\nu}^C$ plus the $\Pi^A_l$ term contribution. Even though that first term would vanish in DR, we keep it to track more easily the UV behaviour in terms of a cutoff $\Lambda\rightarrow\infty$, since in that way one can check the consistency of the different versions of the sum rule. Actually, and this is an important point, the finiteness of the result for $\Lambda\rightarrow\infty$ is directly connected with the well-known Weinberg sum rules \cite{Weinberg:1967kj} (at $T=0$ in the chiral limit):

\begin{eqnarray}
\int_0^\infty  ds \left[ \rho^V(s)-\rho^A(s)\right]=F_\pi^2\label{wsrT0chlim1}\\
\int_0^\infty ds \ s \left[ \rho^V(s)-\rho^A(s)\right]=0  \label{wsrT0chlim2}
\end{eqnarray}

Hence, consider the dominant quadratic $\int d^4q (1/q^2)\sim\Lambda^2$ divergent UV part in the $s$ integral in (\ref{sumruleT0SPLprev}), which is given just by the leading order in the expansion $q^2\gg s$ (formally $Q^2\gg s$ after Wick rotating the integral so that the Minkowskian $q_0\rightarrow -i q_0$ and $Q^2=q_0^2+\vert\vec{q}\vert^2$). That leading contribution cancels then exactly with the first term inside the brackets if (\ref{wsrT0chlim1}) holds. The next to leading UV divergence is of order $\int d^4q (1/q^4)\sim\log\Lambda$  and cancels also once (\ref{wsrT0chlim2}) is used.  Once $\left.\Delta M_\pi^2\right\vert_{SPL, T=0}$ in (\ref{sumruleT0SPLprev}) is shown to be finite, the $Q^2$ Euclidean integral can be performed, giving rise to the original sum rule in \cite{Das:1967it}:

\begin{equation}
\left.\Delta M_\pi^2\right\vert_{SPL, T=0}=-\frac{3e^2}{16\pi^2 F_\pi^2}\int_0^\infty ds s (\ln s) \left[\rho_V(s)-\rho_A(s)\right]
\label{sumruleT0SPL}
\end{equation}

Thus, at $T=0$ and in the chiral limit one gets the typical $\rho_V-\rho_A$ contribution which naively would vanish if chiral symmetry is restored.

For practical purposes, it would be useful to assume that the vector and axial spectral functions are saturated, respectively, by the $\rho$(770) and $a_1$(1260) resonances, consistently  with Vector Meson Dominance and Resonance Saturation (RS) \cite{Ecker:1988te,Ecker:1989yg,Donoghue:1996zn}. See also our comments in section \ref{sec:expres}. In this section, this will only used for power counting arguments regarding the sum rule, rather than to get a numerically accurate prediction. In addition, at least for a rough estimate, we can in principle neglect the width of those resonances with respect to their mass, so that at zero temperature $\rho_{V,A}\sim F_{V,A}^2 \delta \left(s-M_{V,A}^2\right)$ where $F_V^2$ and $F_A^2$ are the constant residues of the current correlators. They correspond respectively to $\rho\gamma\pi^{2n}$  and $a_1\gamma\pi^{2n+1}$ ($n\geq 0$) couplings in the spin-1 resonance lagrangian \cite{Ecker:1989yg}. Recall that, in that limit, (\ref{wsrT0chlim1}) and (\ref{wsrT0chlim2}) would give respectively  $F_V^2-F_A^2=F_\pi^2$ and $F_V^2M_V^2=F_A^2 M_A^2$, which are reasonably fulfilled by the physical values of those constants \cite{Donoghue:1996zn,Ecker:1988te}. When this narrow RS limit is used in (\ref{sumruleT0SPL}), one gets $\left.\Delta M_\pi^2\right\vert_{SPL, T=0}\simeq \frac{3e^2 F_V^2 M_V^2}{16\pi^2 F_\pi^2}  \log (M_A^2/M_V^2)$ which gives $M_{\pi^\pm}-M_{\pi^0}\simeq$ 4.7 MeV,  reasonably close to the experimental value of 4.594$\pm$0.001 MeV.

In general, the vector and axial-vector spectral functions should be more elaborated, including nonzero widths, continuum and excited states contributions, in order to comply with phenomenology data such as $\tau$-decay data (see \cite{Hohler:2012xd} for  a recent update). This level of precision will not be necessary for our present work.

An important point in our analysis will be to classify the different contributions to the pion mass difference according to a power counting in terms of typical resonance masses. Thus, we consider a formal expansion parameter:

 $$x\sim M_\pi^2/M_R^2\sim T^2/M_R^2$$
  where $M_R=\Od(M_{V,A})$. $F_{V,A}$ and $F_\pi$ are treated as parameters of the same order in this expansion. Note that we treat the pion mass and the temperature as being of the same order, which is the main difference of the present work with \cite{Manuel:1998sy}. This counting is basically equivalent to the chiral expansion. However, working within the framework of resonance models will help better to understand the modifications to the sum rule (\ref{sumruleT0SPL}) as well as to make numerical estimates of the accuracy of ChPT, which will be carried out in section \ref{sec:expres}.

Thus, according to our previous discussion,  we can think of the SPL result (\ref{sumruleT0SPL}) as the leading $\Od(M_R^2)$ order, which actually gives the numerically dominant contribution to the constant $C$ in (\ref{treemassessu2}), whereas NLO  corrections  of $\Od(x M_R^2)\sim \Od(M_\pi^2, T^2)$ arise from the ChPT pion loops discussed in section \ref{sec:chpt}.

\subsection{$T\neq 0$ sum rule in the SPL}

Let us now still keep the SPL (and therefore the chiral limit) but allow $T\neq 0$, as in the analysis performed in \cite{Manuel:1998sy}. One can then  assume that the soft pion and current algebra theorems relating the pion expectation value of (\ref{compton}) with current correlation functions, as in (\ref{comptonspl}) still holds. However, a crucial point is that now $\Pi^{V,A}_{\mu\nu}(q;T)$ are  $T$-dependent correlation functions corresponding to $\langle {\cal T} V_\mu^3 (x) V_\nu^3 (0)\rangle_T$ and $\langle {\cal T} A_\mu^3 (x) A_\nu^3 (0)\rangle_T$.

Those  correlators, apart from carrying on $T$-dependent corrections to the spectral functions, will give rise to a more complicated tensor structure, as discussed in Appendix \ref{app:genprop}.  Thus, the steps leading to the thermal version of  (\ref{comptonspl}) are only valid in the SPL  and up to $\Od(T^2)$ corrections. For instance,  $F_\pi^2 (T)$ defined through the residue of the axial correlator at the pion pole gives rise to two independent pion decay constants,  corresponding to the space and time components of the axial current, from $\Od(T^4)$ onwards \cite{Pisarski:1996mt}, even in the chiral limit.

Keeping only the leading $\Od(T^2)$ corrections in the chiral limit, it is well known that the only thermal correction to axial and vector spectral functions is a multiplicative renormalization with respect to the $T=0$ ones, namely \cite{Dey:1990ba}:

\begin{eqnarray}
\Pi^V_{\mu\nu} (q;T)&=&\left[1-\epsilon(T)\right]\Pi^V_{\mu\nu} (q;0)+\epsilon(T) \Pi^A_{\mu\nu} (q;0) \nonumber\\
\Pi^A_{\mu\nu} (q;T)&=&\left[1-\epsilon(T)\right]\Pi^A_{\mu\nu} (q;0)+\epsilon(T) \Pi^V_{\mu\nu} (q;0)
\label{mixingchiral}
\end{eqnarray}
where $\epsilon(T)=\frac{T^2}{6F^2}=2g_1(0,T)/F^2$ comes from pion tadpole corrections. Note that this SPL mixing predicts chiral restoration, in the sense of axial-vector current degeneration, at $\epsilon=1/2$, i.e, at $T\simeq \sqrt{3}F$, before the  value for which the quark condensate   vanishes in the chiral limit, $T\simeq \sqrt{8}F$ \cite{Gasser:1986vb}. Thus, to this order the only modification is the residue of the correlators, not their poles.  Actually, the temperature corrections to the $\rho$,$a_1$ meson masses and widths are expected to be of order $\Od(e^{-M_R/T})=\Od(e^{-1/\sqrt{x}})$ \cite{Rapp:1999ej,Song:1996dg,Rapp:1999qu,Dobado:2002xf}.

Therefore, using in (\ref{compton}), the thermal version of (\ref{comptonspl}) with $F_\pi^2\rightarrow F_\pi^2(T)$ and the VA correlators replaced by (\ref{mixingchiral}), we can write now:

\begin{eqnarray}
\left.\Delta M_\pi^2\right\vert_{SPL, T\neq 0}=4e^2T\sum_n \int \frac{d^3\vec{q}}{(2\pi)^3}\frac{1}{\omega_n^2+\vert\vec{q}\vert^2} - \frac{e^2 \left[1-2\epsilon(T)\right]F_\pi^2}{F_\pi^2 (T)} T\sum_n \int \frac{d^3\vec{q}}{(2\pi)^3}\frac{1}{\omega_n^2+\vert\vec{q}\vert^2}\nonumber\\
-\frac{3e^2 \left[1-2\epsilon(T)\right]}{F_\pi^2(T)}T\sum_n \int \frac{d^3\vec{q}}{(2\pi)^3} \int_0^\infty ds \frac{\rho^V(s;0)-\rho^A(s;0)}{\omega_n^2+\vert\vec{q}\vert^2-s+i\epsilon}
\label{thermalspl1}
\end{eqnarray}
where $F_\pi^2 (T)=F^2\left[1-\epsilon(T)\right]$ in the chiral limit \cite{Gasser:1986vb}.

The first term above comes from the contact term (photon tadpole in Fig.\ref{fig:diag}) and is  the  Debye screening mass of the longitudinal photons which will contribute also to the charged thermal mass. In DR, from (\ref{Gfun}),
 is given by $4e^2g_1(0,T)=\frac{e^2T^2}{3}$, the $T=0$ term  vanishing identically for a massless particle (in this case the photon) as discussed already in section \ref{sec:real}.

The second term in (\ref{thermalspl1}) comes from the $\Pi_l^A (q^2;T=0)$ part of the $T=0$ axial correlator when using the mixing (\ref{mixingchiral}). Note that to the order $T^2$ that we are keeping,  in the SPL the first and second term in (\ref{thermalspl1}) add together giving a net $T^2$ contribution.

Finally, the last term in (\ref{thermalspl1}) is the reminder of the $V-A$ correlator coming from the non-contact part. Now the relevant $T^2$ contribution arises from  the multiplicative factor in front of the integrals. The rest of the thermal contributions coming from that term are the result of evaluating the Matsubara sum, and are essentially of the order of $g_1(M_R;T)\sim e^{-M_R/T}$ if the spectral functions are taken as saturated by the vector and axial-vector lightest resonances, i.e, those contributions are exponentially suppressed in the $x$ counting that we have introduced in section \ref{sec:T0sumrule}.

Note also that the formal expression (\ref{thermalspl1}) is finite up to $\Od(T^2)$ in the UV cutoff $\Lambda$,  by the same reason than for $T=0$, i.e, using the sum rules (\ref{wsrT0chlim1})-(\ref{wsrT0chlim2}), taking into account that the infinities are contained only in the $T=0$ part of the integrals and that the $\Od(\Lambda^2)$ is formally $\Od(xM_R^2)$ so that when extracting that contribution one should not consider the $\epsilon(T)$ corrections in (\ref{thermalspl1}), which would be of higher order, since we are relying on the mixing (\ref{mixingchiral}) which is valid only up to $\Od(xM_R^2)=\Od(T^2)$. On the other hand, the $\log\Lambda$ is $\Od(M_R^2)$ and then, for that logarithmic divergence those $\epsilon(T)$ corrections have to be kept in both the second and third terms in (\ref{thermalspl1}). Alternatively, before using the mixing (\ref{mixingchiral}), it can be proven that the expression remains finite, since the  Weinberg sum rules hold also at finite temperature by replacing the $s$ integrals of spectral functions by energy ones at fixed spatial momentum \cite{Kapusta:1993hq}, which is the correct representation for the thermal correlators, as discussed in Appendix \ref{app:genprop}. Recall that in \cite{Kapusta:1993hq}, these sum rules are derived for the full axial spectral function, i.e, including the longitudinal part.

Thus, in the chiral and SPL limits and to $\Od(T^2)$, using DR one has:

\begin{equation}
\left.\Delta M_\pi^2\right\vert_{SPL, T\neq 0}=\frac{e^2 T^2}{4} + \frac{2Cf(T)e^2}{F^2}
\label{sumruleTSPL}
\end{equation}
where $f(T)=\left(1-\frac{T^2}{6F^2}\right)$ and $C$ given by the leading order (\ref{sumruleT0SPL}), i.e., $C=\frac{F^2}{2e^2}\left.\Delta M_\pi^2\right\vert_{SPL, T=0}=\Od(M_R^2)$. Recall  that (\ref{sumruleTSPL}) includes the corrections of $\Od(xM_R^2)$ to the leading $\Od(M_R^2)$ order, which in the SPL amount either to $\Od(T^2)$ or $\Od(\epsilon(T)M_R^2)$. Further corrections would include either $\Od(\exp (-1/\sqrt{x}))$ or $\Od(x^2 M_R^2)$, the latter entering proportionally to $T^4$  in the chiral limit. The above result was obtained in \cite{Manuel:1998sy} and gives the same answer as taking the chiral and SPL limits in our general ChPT expression (\ref{selfchpt}) as we have actually shown in section \ref{sec:real}.

It is actually instructive at this point to compare the origin of the different terms from the viewpoint of the role of resonances and possible remnants of the naive chiral-restoring $V-A$ behaviour of the $T=0$ expression (\ref{sumruleT0SPL}). Thus, the first term in (\ref{thermalspl1}), the Debye screening term, is the one coming from diagram (c) in Fig. \ref{fig:diag} as given in (\ref{phtad}). The second term in (\ref{thermalspl1}) is nothing but the chiral limit and SPL version of $\Sigma_{\gamma Ex}$ in (\ref{sigma1pe2})  from diagram (d) in Fig.\ref{fig:diag}. Thus, when setting $\vec{p}=\vec{0}$ and $M=0$, that contribution becomes proportional to the tadpole $T^2$, as discussed in section \ref{sec:real}.  These two terms combine into the $T^2$ first term in the r.h.s of (\ref{sumruleTSPL}), the thermal scalar mass. The remaining bit, i.e, the last term in (\ref{thermalspl1}), proportional to the integrated difference of spectral functions, is a tadpole correction coming from diagrams of type (a) in Fig.\ref{fig:diag}, namely the $-4Ze^2g_1$ term in (\ref{selfchpt}).  This term gives rise to the second contribution in the r.h.s of (\ref{sumruleTSPL}) since in the chiral limit, the additional tadpole contribution in (\ref{selfchpt})  vanishes exactly.

Therefore, the chiral restoration $V-A$ behaviour of the mass difference, driven by the function $f(T)$ in (\ref{sumruleTSPL}), which in principle makes the EM pion mass difference decrease, is compensated now by the increasing behaviour of the combined Debye+Photon exchange first term in (\ref{sumruleTSPL}). The numerical size of these two terms are indeed comparable, and the net result is an almost constant $T$ behaviour which masks then the  chiral restoring. This was already noticed in \cite{Manuel:1998sy}. Our purpose here is to show that this behaviour remains and is even more pronounced for physically realistic pion masses, coming from two different sources: the naive extension of the SPL sum rule using now $M_\pi\neq 0$ thermal functions, plus the $\Od(M_\pi^2)$ deviations from that sum rule.   As discussed above, the chiral limit is nothing but the leading asymptotic term for $T\gg M_\pi$. However, for realistic masses, the  corrections are important and actually their analysis allows to understand better the obtained $T$-dependent behaviour.

\subsection{$T\neq 0$ analysis for nonzero pion masses and momenta}

Most of our previous discussion deals with the SPL $p_\mu\rightarrow 0$ with $p$ the external pion four-momentum. In that limit it is mandatory to take the chiral limit, i.e, massless pions for $e=0$ or vanishing quark masses. However, for realistic temperatures such as those being reached in Heavy Ion  experiments, this is not a good approximation, since $T$ and $M_\pi$ are parameters of the same order, and so they are in the chiral expansion.

If the SPL is abandoned and the quark masses are nonzero, some of the previous arguments in this section have certainly to be revisited. We can start from the general equation (\ref{compton}), from which we separate the connected part of the current correlator. However, for the non-connected part, the relation with the thermal correlators $\Pi_{\mu\nu}^{V,A}$ through the thermal extension of (\ref{comptonspl}) does not necessarily hold for $p_\mu\neq 0$. It is also unclear that the  $V-A$ mixing effect (\ref{mixingchiral}) is also the dominant one when replacing $\epsilon(T)\rightarrow 2g_1(M_\pi;T)/F^2$, as it is often considered \cite{Hohler:2012xd}, since the original mixing theorem \cite{Dey:1990ba} was derived precisely assuming the SPL in the connection between pion expectation values and thermal correlators. Note that this replacement would come just from changing the free pion propagator form the massless case to the massive one.

One could then wonder whether the thermal SPL sum rule could be naively extended just by changing the free pion propagator. One way to see that such sum rule extension does not hold, is to look again at the UV behaviour with a cutoff $\Lambda$. Consider then the extension of (\ref{thermalspl1}) replacing just $(\omega_n^2+\vert\vec{q}\vert^2)\rightarrow (\omega_n^2+\vert\vec{q}\vert^2+M_\pi^2)$ in the second integral, to comply with PCAC at $M_\pi\neq 0$, $\epsilon(T)\rightarrow 2g_1(M_\pi;T)/F^2$ and the finite mass correction to $F_\pi^2(T)$ which is just $F_\pi^2(T)=F_\pi^2(0)\left[1-2g_1(M_\pi;T)/F^2\right]$ \cite{Gasser:1986vb}. Note that $F_\pi^2(0)$ receives now corrections of order $x\sim M_\pi^2/\Lambda_\chi^2$. Taking now the leading UV terms, as we did in section \ref{sec:T0sumrule}, the infinities do not cancel, since the WSR (\ref{wsrT0chlim1})-(\ref{wsrT0chlim2}) are known to receive $\Od(M_\pi^2)$ corrections. In particular, (\ref{wsrT0chlim1}) remains the same, but (\ref{wsrT0chlim2}) changes to \cite{Pascual-Narison,Hohler:2012xd}:

\begin{eqnarray}
\int_0^\infty ds \ s \left[ \rho^V(s)-\rho^A(s)\right]=F_\pi^2 M_\pi^2  \label{wsrT0mass2}
\end{eqnarray}

Thus, the leading UV $\Lambda^2$ term corresponding to take $s=0$ in the denominator would still cancel with the Debye term, by the same reasons as discussed in the massless case in the previous section.  Note that for this leading term it is irrelevant to put $M_\pi\neq 0$ in the propagator inside the second integral. However, when the NLO $\log\Lambda$  is considered, there is no cancelation, since the last integral gives an extra factor of 3 when using (\ref{wsrT0mass2}), with respect to the $M_\pi^2$ term in the expansion of the second integral.
Thus, we expect additional $\Od(M_\pi^2)$ and $\Od(\modp^2)$ corrections. Actually, as we did in the chiral and SPL limits, we can read off the full result for the pion mass difference up to order $\Od(e^2xM_R^2)$ from our previous ChPT analysis in section \ref{sec:chpt} since this has to be the model-independent answer to that order.  However, the sum rule analysis presented here will still be useful to keep track of the fate of the chiral-restoring terms, associated to the $V-A$ spectral function differences in the thermal bath, and of the main differences with the chiral limit.

Thus, let us consider the different thermal contributions to the mass difference obtained in our previous ChPT analysis, now for $M_\pi\neq0$.  The Debye term of diagram (c) in Fig.\ref{fig:diag} is given in (\ref{phtad}) and is directly identified with the $T_{\mu\nu}^{(C)}$ contact term as in the SPL/chiral limit.  The remaining  contributions are of three different types, which we discuss in connection with our analysis in this section:

\begin{enumerate}
\item The term with no $F^2$ dependence, namely the pion-photon exchange contribution $\Sigma_{\gamma Ex}(\modp;T)$ given in (\ref{sigma1pe}) and (\ref{sigma1pe2}), which comes from the photon exchange diagram (d) in Fig.\ref{fig:diag}. We can think of this term as the proper extension of the second contribution in (\ref{thermalspl1}) which, apart from the modification of the pion propagator to the massive case, includes the insertion of $\displaystyle \frac{(2p-q)^2}{(p-q)^2+M_\pi^2}$, which takes into account that the pion-photon vertex also receives $p$ corrections. Now this term is not simply proportional to $T^2$ as in the SPL. As discussed in section \ref{sec:real}, its on-shell contribution splits as indicated in eq.(\ref{sigma1pe2}),  giving rise to a $T^2$ term which adds to the Debye one to give the positive $T^2$ term in (\ref{selfchpt}) plus the $e^2 g_1$ and $4M_\pi^2\re J_T$ terms in (\ref{selfchpt}), which are both increasing functions of $T$, as it is clear from the discussion in Appendix \ref{app:loop}.

\item The $-4Ze^2g_1$  term in (\ref{selfchpt}), which comes from  tadpoles (a) in Fig.\ref{fig:diag} and  is therefore proportional to the leading-order EM mass difference as $-2g_1(M_\pi,T)\Delta M_\pi^2/F^2$. Therefore, this term gives the direct $M_\pi\neq0$ extension of the $T^2$ term in $f(T)$ in (\ref{sumruleTSPL}) and thus inherits the $V-A$ chiral restoring behaviour.

\item The remaining  term, i.e, the second one in   (\ref{selfchpt}), coming also from   tadpoles (a) in Fig.\ref{fig:diag}. It has no counterpart in the SPL and therefore it is an $\Od(M_\pi^2)$ modification of the SPL sum rule that has to be taken into account also to this order. Recall that, as indicated in section \ref{sec:real}, this  term can be written,  to $\Od(e^2)$ and $\Od(xM_R^2)$ as $\frac{M_\pi^2}{F^2}\Delta M_\pi^2 g_2(M_\pi,T)+\Od(e^4,x^2M_R^2)$ with $g_2$ in (\ref{g2}).

\end{enumerate}

With the above structure, let us consider again the formal cutoff $\Lambda$ dependence in order to arrive to a  consistent modification of the thermal sum rule. For that purpose, recall  the large $q^2$ expansion  of the $T=0$ part (which contains the UV divergences) of the pion-photon exchange contribution:

\begin{equation}
\int d^4 q \frac{1}{q^2}\frac{(2p-q)^2}{(p-q)^2-M_\pi^2}=\int d^4 q \frac{1}{q^2} \left[1+\frac{4M_\pi^2}{q^2}-2\frac{p\cdot q}{q^2}-4\frac{(p\cdot q)^2}{\left(q^2\right)^2}+\Od(q^{-3})\right]
\label{phexdiv}
\end{equation}
where the on-shell condition $p^2=M_\pi^2$ has been used. Now, taking into account that, at $T=0$:

$$
\int d^4 q \frac{1}{q^2} \frac{p\cdot q}{q^2} =0
$$
by parity, and

$$
\int d^4 q \frac{1}{q^2} \frac{(p\cdot q)^2}{\left(q^2\right)^2}=p_\mu p_\nu \int d^4 q \frac{1}{q^2} \frac{q^\mu q^\nu}{\left(q^2\right)^2}=\frac{1}{4}p_\mu p_\nu g^{\mu\nu} \int d^4 q \frac{1}{q^2} \frac{q^2}{\left(q^2\right)^2}=
\frac{M_\pi^2}{4}\int d^4 q\frac{1}{\left(q^2\right)^2}
$$

we find that  the $\log\Lambda$ contribution in the photon exchange term (\ref{phexdiv}) equals

$$
3M_\pi^2 \int d^4 q \frac{1}{\left(q^2\right)^2}
$$
and therefore cancels with the $\log\Lambda$ part of the $\rho_V-\rho_A$ contribution

$$
\frac{3}{F_\pi^2}\int d^4 q \frac{1}{\left(q^2\right)^2} \int_0^\infty ds \ s \left[ \rho^V(s)-\rho^A(s)\right]
$$
when using the corresponding $\Od(M_\pi^2)$ extension (\ref{wsrT0mass2}) of the WSR. Recall that, as we commented before in the SPL case, when considering the $\log\Lambda$ correction one has to keep the $T$-dependent function multiplying both the pion-photon exchange and the $\rho_V-\rho_A$ contributions.

Therefore, at least at the order considered here, we find that the thermal part of the sum rule (\ref{thermalspl1})-(\ref{sumruleTSPL}) can be consistently modified at $M_\pi\neq 0$ by  a) Replacing the sum and integral in the second term in the r.h.s. of (\ref{thermalspl1}) by the pion-photon exchange contribution (\ref{sigma1pe}) and b) Modifying the $T$-dependent function multiplying the $V-A$ vacuum spectral function difference, i.e $f(T)$ in (\ref{sumruleTSPL}), by:

\begin{equation}
f(T)\rightarrow 1-2\frac{g_1(M_\pi,T)}{F^2}+\frac{M_\pi^2}{F^2}g_2(M_\pi,T)
\label{fmod}
\end{equation}

Such modification is consistent with ChPT (model independent) and with the required UV behaviour at this order, i.e up to  $\Od(xM_R^2)$ as explained. We then see, as anticipated in section \ref{sec:chpt}, that the ''chiral restoring" function $f$ is modified by the $T$-increasing term $g_2$ in (\ref{fmod}) which typically for $T\gg M_\pi$ behaves as $TM_\pi$ instead of the $T^2$ decreasing behaviour (restoring) coming from the $g_1$ part, but which for physically realistic masses and temperatures $T\sim M_\pi$ can be of the same numerical order as the restoring term. In addition, the nontrivial modification of the pion-photon exchange introduces a $p$-dependence, a nonzero imaginary part at this order and an additional $T$-increasing term for the real part. Recall that the scalar thermal mass coming from the Debye term plus the chiral limit of pion-photon exchange, is also growing with $T$ against the chiral behaviour,  so that our analysis in this section of the structure of the sum rule shows that introducing $M_\pi\neq0$ corrections amplifies even further this shadowing effect and there is finally no trace of a recognizable chiral-restoring effect in the EM pion mass difference. Put in different words, and as recalled in section \ref{sec:real}, chiral symmetry is ultimately responsible for keeping the EM pion mass difference almost unchanged and softly increasing with $T$.

The analysis we have just shown clarifies the structure of the sum rule and the formal role of the resonance contributions, to the order considered, equivalent to that in our previous ChPT calculation. Our next step will be to explore to what extent we can trust this order for numerically relevant masses and temperatures. For that purpose, we will consider explicitly a  model in which $\rho$ and $a_1$ resonances are coupled explicitly to pion and photon fields, which allows to estimate the typical size of the corrections to our previous ChPT and sum-rule analysis of the EM self-energy difference.

\section{Explicit resonance analysis}
\label{sec:expres}

In order to estimate the size of the corrections to the ChPT $\Od(p^4)$  result for the EM pion self-energy difference and also to contrast the previous sum rule analysis of the role of resonances and chiral restoration, we will consider the self-energy calculation in a model where $\rho$ and $a_1$ resonances are explicitly included in the lagrangian, within the RS approach. In particular, we will take the  resonance lagrangian in \cite{Ecker:1988te} where resonances are coupled to pions in the chiral  lagrangian. Without electromagnetic effects, the resonance couplings  produce $\Od(p^4)$ contributions to the non-EM LEC when those resonances are integrated out, saturating completely those LEC in the RS limit. Actually, we will consider the RS limit for narrow resonances, which is formally well understood in the large-$N_c$ limit \cite{Masjuan:2007ay} since resonance masses are $\Od(1)$ but resonance widths as well as pion loops are $\Od(1/N_c)$ \cite{Gasser:1984gg,largenc}. Actually, we will formally rely on the large-$N_c$ limit to classify the resonance diagrammatic contributions. We shall see  that a consistent matching with ChPT  would require formally higher order diagrams, although RS to leading order  would be enough to estimate the corrections to the ChPT result and hence its validity range.   With EM interactions, resonances contribute already to the $C$ constant in (\ref{L2}), saturating it almost completely \cite{Ecker:1988te,Baur:1995ig}, which is actually what we have discussed in section \ref{sec:T0sumrule} in the context of Das sum rule \cite{Das:1967it}. Therefore, within the RS hypothesis, we start from the lagrangian in (\ref{L2}) with $C=0$ plus the resonance lagrangian in \cite{Ecker:1988te}, whose relevant propagators and vertices can also be found in that paper.   We consider in this model the diagrams contributing to the EM pion mass self-energy difference. Alternatively, as done for instance in \cite{Donoghue:1996zn,Ladisa:1999wx}, one can start from (\ref{compton}) and write down the relevant Compton scattering diagrams. Thus, to leading order in RS and to $\Od(e^2)$, we consider the one-loop diagrams shown in Fig.\ref{fig:diagres} for the  charged-neutral pion self-energy difference, to be added to diagram (c)  in Fig.\ref{fig:diag}. We do not need to consider tadpole contributions  (diagram (a) in Fig.\ref{fig:diag})  since, as we have just explained, the tree level charged and neutral pion masses are the same to leading order in RS. In that sense, note that   pion loops  carry also  additional factors $F^{-2}=\Od(N_c^{-1})$. Diagram (b) in Fig.\ref{fig:diag} has to be considered formally to absorb the loop divergences in the corresponding EM LEC \cite{Baur:1995ig}, which is a $T=0$ contribution not altering our finite $T$ analysis. Actually, by the RS procedure, one finds also the finite resonance contribution to those LEC \cite{Baur:1995ig}.  Note also that diagram (e) in Fig.\ref{fig:diagres} represents the extension of diagram (d) in Fig. \ref{fig:diag} when the $\pi\pi\gamma$ vertex is corrected by a form factor coming from $\rho$ exchange.

Let us then consider the contribution to the neutral-charged self-energy difference of the finite temperature integrals  corresponding to the diagrams in Fig.\ref{fig:diagres}. After some algebraic manipulations, similar to those performed in section \ref{sec:chpt}, we can write them in terms of the $G$ and $J_T$ functions described in Appendix \ref{app:loop} as follows:

\begin{figure}
\centering
\includegraphics[scale=0.17]{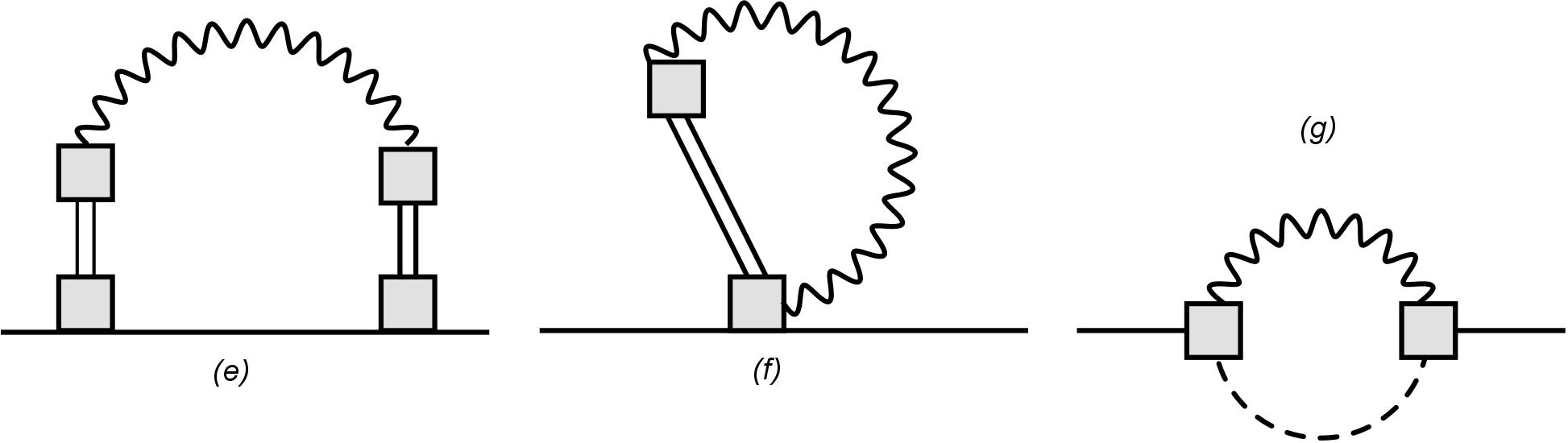}
\caption{Resonance Saturation 1-PI diagrams contributing at leading one-loop order to the charged-neutral pion self-energy  difference.  $\rho$ and a$_1$ particles are represented by double and dashed lines, respectively. The relevant vertices including charged pions, resonances and photons are drawn as grey boxes.}
\label{fig:diagres}
\end{figure}

\begin{eqnarray}
\label{formfactor}
\Delta\Sigma^{(e)}&=&-4e^2M_\rho^4\tintq \frac{(p\cdot q)^2-p^2q^2}{q^2\left(M_\rho^2-q^2\right)^2\left((p-q)^2-M^2_\pi\right)}+e^2\tintq\frac{1}{q^2}
\nonumber\\&=&\Sigma_{\gamma Ex}(\omega+i\epsilon,\omega^2=E_p^2;T)+e^2\left[1-M_\rho^2\frac{\partial}{\partial M_\rho^2}\right]\left[G(M_\rho,T)-G(M_\pi,T)\right.\nonumber\\
&-&\left.(4M_\pi^2-M_\rho^2)J_T(M_\rho,M_\pi;\omega+i\epsilon,\omega^2=E_p^2)\right]
\end{eqnarray}

\begin{equation}
\Delta\Sigma^{(f)}=-3e^2\left(\frac{F_V}{F_\pi}\right)^2\tintq \frac{1}{M^2_\rho-q^2}=-3e^2\left(\frac{F_V}{F_\pi}\right)^2G(M_\rho,T)
\label{rhophotad}
\end{equation}

\begin{eqnarray}
\Delta\Sigma^{(g)}&=&3e^2\left(\frac{F_A}{F_\pi}\right)^2\tintq\frac{1}{M^2_{a_1}-q^2}\nonumber\\&+&2e^2\left(\frac{F_A}{F_\pi}\right)^2\tintq\frac{1}{M_{a_1}^2\left(M_{a_1}^2-(p-q)^2\right)}\left(\frac{(p\cdot q)^2}{q^2}-p^2\right)\nonumber\\
&=&\frac{1}{2}e^2\left(\frac{F_A}{F_\pi M_{a_1}}\right)^2\left[\left(5M_{a_1}^2-M_\pi^2\right)G(M_{a_1},T)+(M_{a_1}^2-M_\pi^2)G(0,T)\right.\nonumber\\
&-&\left.(M_{a_1}^2-M_\pi^2)^2 J_T(0,M_{a_1};\omega+i\epsilon,\omega^2=E_p^2)\right]
\label{a1pholoop}
\end{eqnarray}
with $\Sigma_{\gamma Ex}(\omega+i\epsilon,\omega^2=E_p^2;T)$ the ChPT contribution in eq.(\ref{sigma1pe2}), $E_p^2=\modp^2+M_\pi^2$ and we have taken $F_VG_V=F^2$ in the (e) contribution, where $G_V$ is the coupling constant entering the $\rho\pi\pi$ vertex \cite{Baur:1995ig,Ecker:1989yg}.  The $T=0$ contributions of the above diagrams, which include the UV divergent part to be absorbed in the low-energy constants, can be found in \cite{Baur:1995ig}.

In connection with our discussion in previous chapters, let us discuss  the $x$-expansion (defined in section \ref{sec:T0sumrule}) of the different contributions.  The leading order $\Od(M_R^2)$  to the self-energy difference comes from the $T=0$ part of diagrams (f) and (g) and one can check that its UV $\lambda$-pole contribution in DR cancels precisely using the leading part of the WSR (\ref{wsrT0mass2}), i.e, $F_V^2M_V^2=F_A^2M_A^2+\Od(xM_R^2)$. Recall that within the RS approach, we are taking the resonance spectral functions as completely saturated by the $\rho$ and $a_1$ poles. On the other hand, its finite part gives precisely the narrow resonance limit of (\ref{sumruleT0SPL}), i.e, the value for  $C=\frac{3F_V^2 M_V^2}{32\pi^2}  \log (M_A^2/M_V^2)$ \cite{Das:1967it} in (\ref{treemassessu2}), which saturates the pion mass difference at $T=0$, accordingly with the RS hypothesis and with our discussion in Section \ref{sec:T0sumrule}. In turn, note that the form factor contribution (f) in (\ref{formfactor}) is UV finite as can be checked from direct power counting and by the cancelation of the $\lambda$ pole in the expression given in (\ref{formfactor}) in terms of $G$ and $J_T$, recalling the pole contribution of these two functions given in Appendix \ref{app:loop}.

To $\Od(xM_R^2)$, the form factor contribution (e) in (\ref{formfactor}) reduces to the first term $\Sigma_{\gamma ex}$, which is the ChPT result of diagram (d) analyzed in section \ref{sec:chpt}. We have checked that the remaining terms in (\ref{formfactor}), once their $T=0$ par is separated, do not contribute to this order, expanding the $J_T$ term in inverse powers of $M_{\rho}^2$. On the other hand, diagrams (f) and (g) both contribute with  a zero temperature  $M_\pi^2 \lambda$ pole. In the case of diagram (f), that pole comes from  including the $M_\pi^2$ correction in the WSR, i.e, $F_V^2M_V^2=F_A^2M_A^2 + F_\pi^2M_\pi^2$ according to (\ref{wsrT0mass2}). The $T$-dependent part of (\ref{rhophotad}) is exponentially suppressed as $\Od\left(e^{-M_\rho/T}\right)=\Od\left(e^{-1/\sqrt{x}}\right)$ according to (\ref{g1asymlowT}), while in (\ref{a1pholoop}) we have also checked that the  $\Od(T^2)$ contributions coming from  the $M_{a_1}^2G(0,T)$ and $M_{a_1}^4 J_T$ terms cancel each other, once the $J_T$ is expanded in inverse powers of $M_{a_1}^2$.

An important comment at this point is that one does not recover from the leading order RS approach the full result of the ChPT calculation given in (\ref{selfchpt}). The second  term and the $-4Zg_1$ contribution in the r.h.s of (\ref{selfchpt}), both coming from tadpole diagrams of the type (a) in Fig.\ref{fig:diag}, appear  in higher order diagrams in the RS expansion. For instance, diagrams of type (a) in Fig.\ref{fig:diag} in which one of the internal charged lines is dressed with the resonance diagrams in Fig.\ref{fig:diagres} will contribute to the second term in the r.h.s of (\ref{selfchpt}). Also, diagrams in Fig.\ref{fig:diagres} in which a pion tadpole is attached to the $\rho\gamma\pi\pi$ or to the $a_1\gamma\pi$ vertices, would contribute as $Zg_1$. In addition, vector and axial vector propagators are modified by loop diagrams  beyond RS. Their residues are meant to contribute also at $\Od(T^2)\sim \Od(xM_R^2)$  via the $\rho-a_1$ mixing effect discussed in section \ref{sec:reso} \cite{Dey:1990ba,Song:1995ga,Mallik:2002ef}, while the mass and width modifications of the spectral functions are expected to be of $\Od\left(e^{-1/\sqrt{x}}\right)$\cite{Song:1995ga,Dobado:2002xf,Rapp:1999ej,Song:1996dg}. The $\rho\gamma$ coupling can also receive finite $T$ corrections \cite{Song:1996dg}. Some of those corrections to the EM self-energy difference, but clearly not all of them, could be parametrized in a  $T$-dependent form factor as considered in \cite{Ladisa:1999wx}.

In any case, what is  relevant for our present discussion  is that all these higher order diagrams come with prefactors coming  from the vertices, which are formally subleading in the $1/N_c$ counting, for instance those coming with inverse powers of $F^2$ in (\ref{selfchpt}), as compared to those considered in Fig.\ref{fig:diagres}. This is the formal way to keep track of the leading RS contributions. As emphasized above, we will stick here to the strict RS limit, which is consistent with considering free resonance spectral functions with zero widths, in order to estimate the size of the corrections to the ChPT analysis. Actually, we recall that due to the model independency of the ChPT framework, we are sure that the final answer to $\Od(xM_R^2)$ is that given by (\ref{selfchpt}). Therefore,  we estimate the  corrections as the result of evaluating the thermal contributions (\ref{formfactor})-(\ref{a1pholoop}) once the $T=0$ and the $\Od(xM_R^2)$ given by the first term in the r.h.s of (\ref{formfactor}) are subtracted. In doing so, we note that the next order of correction is actually $\Od(x^2M_R^2)$. In particular, there are $\Od(g_1(M_\pi,T) M_\pi^2/M_\rho^2)$ and $\Od(T^2M_\pi^2/M_{a_1}^2)$ terms  arising, respectively,  from (\ref{formfactor}) and (\ref{a1pholoop}).  Note that these terms are not present in the chiral limit. As commented above, the contribution (\ref{rhophotad}) is exponentially suppressed, and,  we have also checked that  the imaginary part contributions coming from  (\ref{formfactor}) and (\ref{a1pholoop}) are also exponentially suppressed with respect to the ChPT result arising from the first term in (\ref{formfactor}) and analyzed in section \ref{sec:im}.

\begin{figure}
\centering
{\includegraphics[scale=.7]{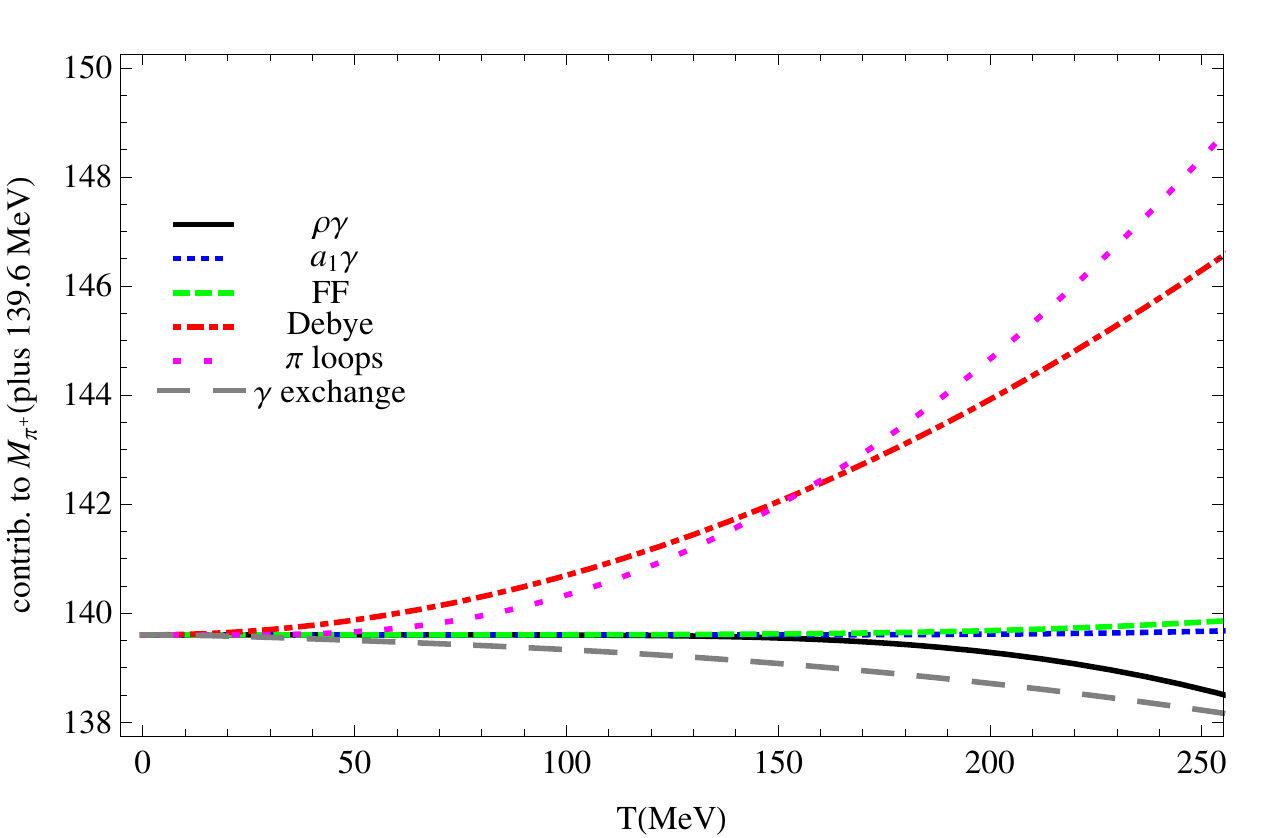}\includegraphics[scale=.75]{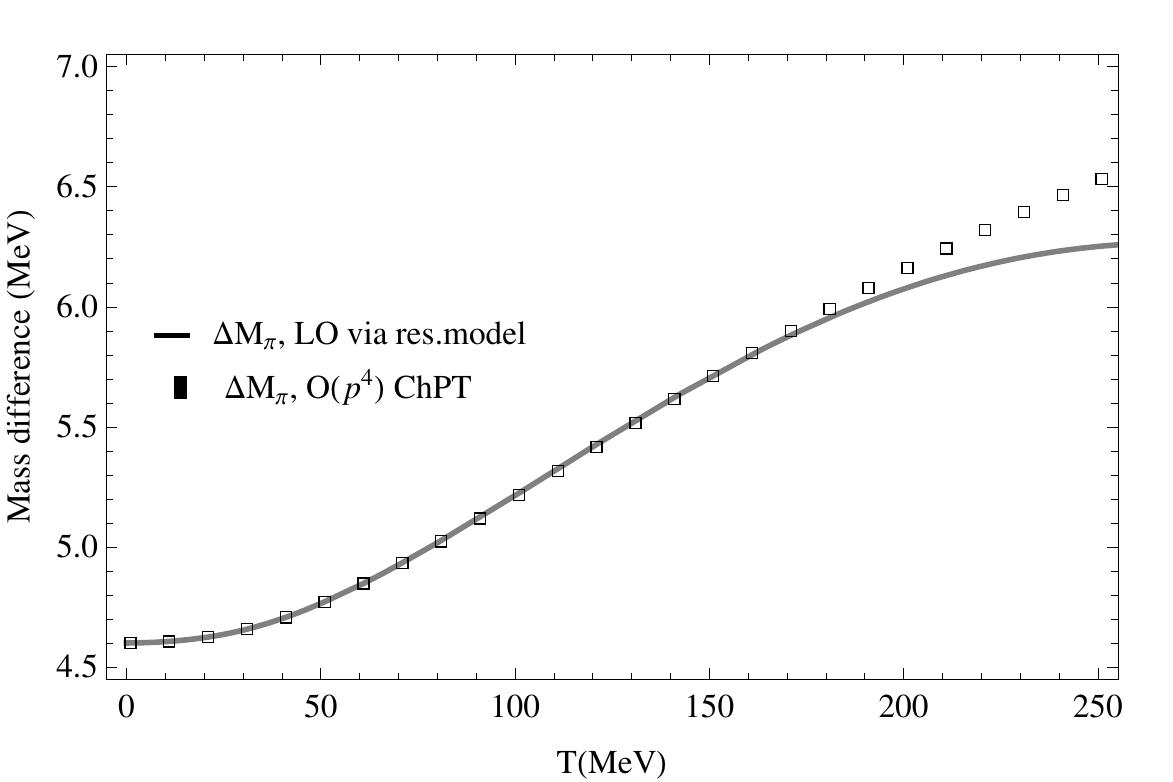}}
\caption{Left: The different thermal contributions to the charged pion mass, including  the resonance model ones, as explained in the main text. Right: Comparison between the EM mass difference obtained with just ChPT and that including the Resonance Saturation leading order contributions. }
\label{fig:contribres}
\end{figure}

We have evaluated numerically the resonance contributions to the real part of the self-energy in the static limit, in order to get an approximate idea of the expected size of the corrections to the ChPT result.  The results are showed in Fig.\ref{fig:contribres}. In the left panel of that figure, we show the different thermal contributions to the charged pion mass given by the diagrams in Figs.\ref{fig:diag} and \ref{fig:diagres}, all shifted to the $T=0$ mass, and discussed here and in section \ref{sec:chpt}. Namely,  the pion tadpole loops given generically by diagram (a) in Fig.\ref{fig:diag}, the Debye term from diagram (c) in Fig.\ref{fig:diag}, the photon exchange term from diagram (d) in Fig.\ref{fig:diag}, the form factor (FF) contribution of diagram (e) in Fig.\ref{fig:diagres} excluding the ChPT photon exchange term, the $\rho\gamma$ photon loop of diagram (f)  in Fig.\ref{fig:diagres} and the $a_1\gamma$ exchange of diagram (g) in Fig.\ref{fig:diagres}. In the right panel we show the deviations of the charged-neutral mass difference calculated within the resonance model with respect to the same ChPT calculation. The numerical values of $F_V$, $F_A$ and $G_V$ are those of \cite{Ecker:1988te}, compatible with $F_VG_V=F^2$. There are no significant changes when using values coming from more recent fits like \cite{Leupold:2003jb,Guo:2011pa}. We have used the physical masses for the resonances, namely $M_{a_1}=1260$ MeV and $M_\rho=770$ MeV.

From this plot, we observe that resonant contributions additional to the ChPT result  activate thermally around 170-200 MeV, which leaves big room for the validity range of the ChPT result. We recall that those resonant contributions do not include $\rho-a_1$ mixing to leading order, which is accounted for already in the ChPT result, as discussed above.   Therefore,  the ChPT calculation for this  observable is dominant and robust  throughout its own applicability range, i.e, below the chiral phase transition. It must be pointed out that, in addition to the size of the absolute value of those resonant corrections, there is an  approximate numerical cancelation between the FF and $\rho\gamma$   terms, as can be seen in the left panel of Fig.\ref{fig:contribres}.

\section{Conclusions}
\label{ref:conc}

In this work we have performed a thorough analysis of the electromagnetic effects in the pion self-energy at finite temperature, within Chiral Perturbation Theory to one loop, which allows to obtain model-independent results, and also including the effect of vector and axial-vector resonant states. The latter have been studied within the context of sum rules and in a explicit resonance saturation approach and allows to understand in a clearer way which contributions come from chiral restoration via $V-A$ mixing. Apart from the link with  chiral symmetry restoration, particular attention has been paid  to discuss phenomenological effects and   gauge invariance.

Within one-loop ChPT we have provided the full expressions for the charged and neutral pion self-energy for physical pion masses and external momenta. There are important  differences with respect to a previous calculation in the chiral limit and vanishing external momentum.  The real part of the self-energy and hence the dispersion relation is momentum dependent. That dependence is rather soft for the relevant range of temperatures, which we have studied by comparing the momentum-averaged self-energy, weighted by pion thermal distributions, with the mass defined in the static limit. Including the physical pion mass gives rise to new terms making the EM pion mass difference increase with temperature. The net result is a  soft increasing behaviour for that difference, which is compatible with it being undetected in the neutral-charged pion spectra observed in Heavy ion Collisions,  with the measurements performed so far. The increasing is softer for the momentum averaged mass than for the static one. The important formal point here is that chiral symmetry restoration via vector-axial vector mixing plays an important role for keeping that difference small, which follows from our combined ChPT and resonance analysis.

Another important conclusion of our present work is the analysis of the EM damping rate for charged pions from the imaginary part of the self-energy. Here it is crucial to work in a physical gauge, we choose the strict Coulomb gauge, to get a meaningful answer, since only physical photon degrees of freedom are in thermal equilibrium. Thus,  the contributions to the imaginary part come from bremsstrahlung-like  processes with physical quasiparticle thermal photons at vanishing spatial momentum, whose contribution is thermally enhanced,  giving rise to an infrared finite result at this order for the imaginary part of the retarded self-energy. The result for the EM damping rate comes only from the transverse modes, it is linearly increasing with temperature, vanishes at zero pion momentum and behaves asymptotically as a constant for large momentum. We have analyzed possible phenomenological consequences of this result. The electromagnetic damping is added to the standard ChPT one so that mean free paths and free times of charged and neutral pions become different. The electromagnetic corrections are comparable in size to the neutral ones up to  $T\sim 60$ MeV. Transport coefficients are expected to be reduced by this effect around a 10\% near the kinetic freeze-out region, with a larger effect in the electrical conductivity than in viscosities. The  freeze-out temperatures for charged and neutral components would also be different, although the expected effect is only about 2 MeV.

We have also studied in detail how the sum rule relating the electromagnetic pion mass difference in the soft and chiral limits with the $V-A$ spectral function difference, is modified by the inclusion of a finite pion mass and nonzero momentum. The standard derivation of the sum rule is no longer applicable and we have found  the required modifications  in order to match the ChPT model-independent result at finite temperature. These are the modification of the photon-exchange contribution to account for the mass and momentum dependence, as well as the multiplicative function in the $V-A$ spectral function difference, which acquires an additional mass-dependent $T$-dependent increasing term. This analysis has been performed to leading  and next to leading order in the expansion in $x\sim T^2/M_R^2 \sim M_\pi^2/M_R^2$ with $M_R$ the resonance masses, i.e, including $\Od(M_R^2)$ and $\Od(xM_R^2)$, equivalent to the ChPT analysis.

In order to confirm the ChPT and sum-rule analysis and also to estimate the next order corrections, we have carried out a explicit calculation of the corrections to the electromagnetic pion self-energy difference at finite temperature within a resonance saturation approach. Thus, we have been able to estimate next to next to leading order corrections, which show up at  $\Od(x^2M_R^2)$. Those corrections remain numerically  small  for the range of temperatures relevant within Heavy Ion Collisions, which results in a rather large applicability range of our ChPT analysis.

\section*{Acknowledgments}
Work partially supported by the Spanish Research contract FPA2011-27853-C02-02 and the FPI programme (BES-2009-013672). We acknowledge the support
of the EU FP7 HadronPhysics3 project.

\appendix

\section{General definitions and properties of spectral functions and dispersion relations}
\label{app:genprop}

Throughout this work we follow closely \cite{galekapustabook} and \cite{lebellac} regarding the finite-temperature formalism. We summarize in these Appendices the most relevant results for our purposes in this work.

For a scalar field or current, the time-ordered version of the propagator in the Euclidean IT Formalism is given by:

$$
G(\vec{x},\tau)=\langle {\cal T} \phi (\vec{x},-i\tau) \phi (0)\rangle_T
$$
where the subscript $T$ indicates a thermal average, $\omega_n=2\pi n T$ is the bosonic Matsubara frequency with $n\in\IZ$ and time-ordering ${\cal T}$ is along $t=-i\tau$ with $\tau\in[-\beta,\beta]$ (time differences). Its Fourier representation can be written as:

\begin{equation}
G(i\omega_n,\modp)=\int_T d^4x  G(\vec{x},\tau) e^{-i\omega_n \tau} e^{-i\vec{p}\cdot \vec{x}}=\frac{1}{\omega_n^2+E_p^2+\Sigma (i\omega_n,\modp;T)}
\label{fullpropit}
\end{equation}
where $\int_T d^4x\equiv\int_0^\beta d\tau \int d^3\vec{x}$, $E_p^2=\modp^2+M_0^2$ and $M_0^2$ is the tree level mass. We will keep the $(+,-,-,-)$ metric with the Euclidean $p_0^E\equiv i\omega_n$  so that we write for instance $p^2=(i\omega_n)^2-E_p^2$ which will become the Minkowski $p^2$ after analytic continuation (see below). In the above equation, $\Sigma$ is the IT self-energy function, which in the thermal case depends independently  on frequency and three-momentum and  explicitly on  $T$.

The analytical continuation from external discrete frequencies to continuous ones can be carried out once all the internal Matsubara sums have been performed and gives rise to the retarded and advanced propagators defined as:

\begin{equation}
G_{R,A} (\omega,\modp)=\mp i G (i\omega_n=\omega\pm i\epsilon,\modp)
\label{retadv}
\end{equation}
with $\omega\in\IR$ and $\epsilon>0$ and we define from these propagators the spectral function as   $\rho(\omega,\modp)=2\im iG_R(\omega,\modp;T)$ whose main properties we discuss below. The spectral function is odd in $\omega$ and in the free case, for which $\Sigma=0$, it reads $\rho_0(\omega,\modp)=2\pi\sgn(\omega)\delta(\omega^2-E_p^2)$.

In the interacting case and in the perturbative regime considered in this paper (see comments below), the self-energy contributions come from loop diagrams  which generate cuts for $\im\Sigma$ along the real axis, so that we write  $\im\Sigma (\omega\pm i\epsilon,\modp)=\mp2\omega \Gamma(\omega,\modp)$ with $\Gamma >0$ along the cuts.

The dispersion relation is determined by the poles of $G_R(\omega,\modp)$, which lie  below the real axis, or equivalently by  the spectral function. If we denote the position of the poles by  $z_{pole}=\omega_p-i\gamma_p$,
with $\gamma_p>0$ the thermal damping rate, we have then $z_{pole}^2=E_p^2+\re\Sigma(z_{pole},\modp;T)-2iz_{pole}\Gamma(z_{pole},\modp;T)$.

In this work we will work within the perturbative regime: $\Sigma\ll E_p^2$, $\omega_p^2=E_p^2(1+\Od(\Sigma/E_p^2))$, $\Gamma_p=\Od(\Sigma/E_p)$ so that the perturbative solution of the pole equations reads $\omega_p^2=E_p^2+\re\Sigma(E_p,\modp;T)$, $\gamma_p=\Gamma(E_p,\modp;T)=-\im\Sigma(E_p+i\epsilon,\modp)/2E_p$, where we have made use of the fact that $\re\Sigma (\omega,\modp)$ and $\Gamma(\omega,\modp)$ are  even functions of $\omega$. Thus, there are two perturbative poles at $\pm \omega_p-i\gamma_p$.

From the previous properties, one can define a complex function $G(z,\modp)$ for complex $z$ analytic for $z$ off the real axis and such that the IT propagator is $G(z=i\omega_n,\modp)$ and  the retarded/advanced propagators are  $G_{R,A} (\omega,\modp)=\mp i G(z=\omega\pm i\epsilon,\modp)$ with $\omega\in\IR$, i,e,

 \begin{equation}G(z,\modp)=\frac{-1}{z^2-E_p^2-\Sigma(z,\modp;T)}\label{fullpropz}\end{equation}

In particular, in the perturbative regime described above, it is easy to check that the above function does not have (perturbative) poles and has the same cuts as $\Sigma(\omega)$ along the real axis.

Let us comment also on the spectral function representation of the different propagators. Applying Cauchy's theorem to $G(z)$ in (\ref{fullpropz}), with its analytical structure discussed above, on a suitable contour  surrounding the real axis from above and from below, one arrives to a dispersion relation valid for the retarde/advanced propagators  and for the IT one, from the same spectral function, namely:

\begin{equation}
G(z,\modp)=\int_{\infty}^\infty \frac{d\omega'}{2\pi}\frac{\rho(\omega',\modp;T)}{\omega'-z} \qquad (z\not\in\IR)
\label{spectralomega}
\end{equation}
Thus, $z=i\omega_n$ correspond to the IT propagator and $z=\omega\pm i\epsilon$ to the retarded/advanced ones.

The above frequency representation is the more adequate one when working at finite temperature. As commented, the analytical continuation of the IT propagator yields naturally the retarded propagator, which has the correct analytic structure in terms of the physical states. In addition, it is valid for any cut structure of $G$ along the real axis, including possible Landau-like purely thermal cuts (see below). It is possible also to define thermal expectation values of $T$-ordered products along $t\in\IR$, within the so-called real-time formalism of Thermal Field Theory \cite{lebellac}. However, those real-time $T$-ordered products do not have a  representation like (\ref{spectralomega}), not even in the free case, nor they describe the spectral properties of the theory in the general interacting case. The problem of how to obtain the retarded correlator from the RT one is discussed in \cite{Kobes:1990kr}.

It is instructive  to relate the above ''energy" spectral representation with the usual $s$-representation used customarily at $T=0$. First, let us write (\ref{spectralomega}) as:

\begin{equation}
G(z,\modp)=\int_{0}^\infty \frac{d\omega'}{2\pi}\frac{2\omega'\rho(\omega',\modp;T)}{(\omega')^2-z^2} \qquad (z\not\in\IR)
\label{spectralomegapos}
\end{equation}
Now, denoting $s=z^2-\modp^2$ and $s'=(\omega')^2-\modp^2$ and assuming that the following two conditions hold: i) $\rho(\omega'>0,\modp)$ is a function only of $s'$, so that $G$ is only a function of $s$ and ii)  $G(s)$ is analytic for $0>s\in\IR$, the lower limit of integration in (\ref{spectralomegapos}) can be extended to $-\modp$ so that by changing variables from $\omega'$ to $s'$ on ends up at $T=0$ with:

\begin{equation}
G(s)=\int_0^\infty ds' \frac{\hat \rho (s')}{s-s'} \qquad (s\not\in \IR)
\label{spectralsT0}
\end{equation}
with $\hat\rho(s')=(-1/\pi)\im G(s'+i\epsilon)$ for $\omega'>0$. Note also the  $2\pi$ factor conventionally included in the normalization of the spectral function at $T=0$.  Alternatively, one can arrive to (\ref{spectralsT0}) directly from the analytic properties of $G$ in the $s$ complex plane. It is important to remark  that none of the conditions i) and ii) above are met at $T\neq 0$ since Lorentz covariance is broken and Landau cuts may be present. At $T=0$, the representation (\ref{spectralsT0}) allows to define the $T$-ordered product $-iG(s+i\epsilon)$ with $s\in\IR$.

For the case of conserved vector and axial-vector current propagators at finite temperature, there are two independent tensor structures $P_T^{\mu\nu}$, $P_L^{\mu\nu}$ which are four-dimensionally transverse \cite{galekapustabook}, $P_T$ being also three-dimensional transverse:

\begin{eqnarray}
P_T^{ij} (q)&=& \delta^{ij}-\frac{q^i q^j}{\modq^2} ; \qquad P_T^{00}=P_T^{0i}=P_T^{i0}=0 \nonumber\\
P_L^{\mu\nu} (q)&=&\frac{q^\mu q^\nu}{q^2}-g^{\mu\nu}-P_T^{\mu\nu}
\label{Tprojectors}
\end{eqnarray}
 where $q^0=i\omega_n$. We remind that the metric signature here is $(+---)$.

 Therefore, any correlator of conserved vector or axial-vector currents can be written as

 \begin{equation}
 \Pi^{\mu\nu}(i\omega_n,\vec{q})=\Pi^T (q) P_T^{\mu\nu} (q) + \Pi^L (q) P_L^{\mu\nu} (q)
\label{LTvector}
\end{equation}

At $T=0$, one has simply $\Pi^T=\Pi^L\equiv\Pi$ so that $\Pi^{\mu\nu}(q)=\left(\frac{q^\mu q^\nu}{q^2}-g^{\mu\nu}\right)\Pi (q)$.
\footnote{Our convention for vector and axial-vector current correlators corresponds to that in \cite{Das:1967it,Kapusta:1993hq,Holt:2012wr,Hohler:2012xd} but differs from \cite{Donoghue:1996zn,Manuel:1998sy}. The latter authors include an additional $q^2$ multiplying the $\Pi (q)$ functions.}

For the photon case, its Euclidean propagator in an arbitrary covariant gauge reads:

$$
D^{\mu\nu}(i\omega_n,\vec{q})=\frac{1}{\omega_n^2+\vert\vec{q}\vert^2+\Sigma_T(i\omega_n,\vec{q})}P_T^{\mu\nu} (q)+
\frac{1}{\omega_n^2+\vert\vec{q}\vert^2+\Sigma_L(i\omega_n,\vec{q})}P_L^{\mu\nu} (q)+\alpha\frac{q^\mu q^\nu}{(q^2)^2}
$$

$$\Sigma^{\mu\nu}=\Sigma_T P_T^{\mu\nu} + \Sigma_L P_L^{\mu\nu}$$

so that the free ($\Sigma=0$) Euclidean photon propagator  is:

\begin{equation}D_0^{\mu\nu}(i\omega_n,\vec{q})=\frac{g^{\mu\nu}}{q^2}+(\alpha-1) \frac{q^\mu q^\nu}{(q^2)^2}
\label{photpropcov}
\end{equation}

As explained in the text, we will also need the free photon propagator in the strict ($\alpha=0$) Coulomb gauge, which reads \cite{galekapustabook}:

\begin{equation}
D_0^{\mu\nu}(i\omega_n,\vec{q})=-\frac{g^{\mu 0} g^{\nu 0}}{\modq^2}-\frac{P^{\mu\nu}_T}{q^2}
\label{photpropcoulomb}
\end{equation}

\section{Thermal loop functions for self-energies}
\label{app:loop}

We describe here the main properties of the typical thermal loop integrals appearing throughout this work. They come from the corresponding $T=0$ one through the replacements
\begin{equation}q_0\rightarrow i\omega_n=i 2\pi n T \quad , \quad  \int\frac{d^4 p}{(2\pi)^4}\rightarrow i T\sum_n\int\,\frac{d^3 \vec{p}}{(2\pi)^3}\label{itrep}\end{equation}
 in the IT formalism, with $n\in\IZ$.

First, consider the tadpole integral of the free propagator:

\begin{eqnarray}
G(M,T)=T\sum_{n=-\infty}^\infty\qint \frac{1}{\omega_n^2+E_q^2}=G(M,0)+g_1(M,T)
\label{Gfun}
\end{eqnarray}
with:

\begin{equation}
g_1(M,T)=\frac{1}{2\pi^2}\int_0^\infty dq \frac{q^2}{E_q} n_B(E_q),
\label{g1}
\end{equation}
with $E_q\equiv\sqrt{q^2+M^2}$,

\begin{equation} n_B(x)=\frac{1}{e^{\beta x}-1}
\label{bose}
\end{equation}
and the  $T=0$ part containing the UV divergence ($T\neq 0$ UV divergences are always contained in the $T=0$ part) is given in dimensional regularization  $D=4-\epsilon$ by:

\begin{equation}
G(M,0)=
2M^2\lambda+\frac{M^2}{16\pi^2}\log\frac{M^2}{\mu_\chi^2}
\label{Gfunsep}
\end{equation}
with
\begin{equation}\lambda=\frac{1}{2}(4\pi)^{-D/2}\Gamma\left(1-\frac{D}{2}\right)\mu_\chi^{D-4}\label{lambdadr}\end{equation}
being $\mu_\chi$ the renormalization ChPT scale and $\Gamma$ the Euler gamma function. For the $T=0$ part we follow the same notation as in \cite{Gasser:1983yg,Gasser:1984gg}.

The $g_1(M,T)$ function has the following asymptotic behaviours:

\begin{eqnarray}
T\gg M: \ g_1(M,T)&=&\frac{T^2}{12}\left[1-6\frac{M}{T}+\Od(\frac{M^2}{T^2}\log\frac{M}{T})\right]\label{g1asymhighT}\\
T\ll M: \ g_1(M,T)&=&(2\pi)^{-3/2}\left(\frac{M}{T}\right)^{1/2}e^{-M/T}\left[1+\Od(T/M)\right]+\Od(e^{-2M/T})
\label{g1asymlowT}
\end{eqnarray}

Second, we analyze the one-loop integral appearing in self-energy diagrams:

\begin{equation}
J_T(m_1,m_2;i\omega_m,\modp)=T\sum_{n=-\infty}^\infty\qint \frac{1}{q^2-m_1^2}\frac{1}{(q-p)^2-m_2^2}
\label{JTm1m2}
\end{equation}
for arbitrary masses $m_1$ and $m_2$.

As discussed above, we are interested in the  analytic continuation of the above integral $i\omega_m\rightarrow z$ for complex $z$ off the real axis. In particular for $z=\omega+i\epsilon$ with $\omega\in\IR$, that would give rise to the retarded function appearing in the retarded self-energy and hence describing the dispersion relation as explained in Appendix \ref{app:genprop}. The analytic continuation is performed after evaluating the internal Matsubara sum in $n$, which can be carried out using standard finite-temperature methods. In fact, inserting the spectral representation (\ref{spectralomega}) for the two IT propagators inside the integral and using the formula:

\begin{equation}
T\sum_n\frac{1}{\omega_1-i\omega_n}\frac{1}{\omega_2-i(\omega_m-\omega_n)}=\frac{n_B(\omega_1)-n_B(-\omega_2)}{\omega_1+\omega_2-i\omega_m}
\end{equation}
we arrive to the retarded continuation of $J_T$:

\begin{eqnarray}
J_T(m_1,m_2;z,\modp)&=&-\qint \frac{1}{4E_{1}E_{2}}\left\{\left[1+n_B(E_{1})+n_B(E_{2})\right]\left[\frac{1}{z-E_1-E_2}-\frac{1}{z+E_1+E_2}\right]\right.
\nonumber\\&+&\left.\left[n_B(E_{1})-n_B(E_{2})\right]\left[\frac{1}{z+E_1-E_2}-\frac{1}{z-E_1+E_2}\right]\right\}
\label{Jcomplex}
\end{eqnarray}
where we have used $n_B(x)+n_B(-x)+1=0$ and, for simplicity, we denote $E_{1}=\sqrt{\vert\vec{q}\vert^2+m_1^2}$, $E_{2}=\sqrt{\vert\vec{q}-\vec{p}\vert^2+m_2^2}$.

Thus, setting $z=\omega+i\epsilon$ with $\omega\in\IR$ and separating the real and imaginary parts gives:

\begin{eqnarray}
\label{reJ}
\re J_T(m_1,m_2;\omega,\modp) &=& \re J_{T=0}(m_1,m_2;\omega,\modp) -\frac{1}{2}{\cal P}\qint\left\{
\frac{n_B(E_1)}{E_1}\left[\frac{1}{(E_1-\omega)^2-E_2^2}+\frac{1}{(E_1+\omega)^2-E_2^2}\right]
\right.\nonumber\\ &+&\frac{n_B(E_2)}{E_2}\left.\left[\frac{1}{(E_2-\omega)^2-E_1^2}+\frac{1}{(E_2+\omega)^2-E_1^2}\right]\right\}\\
\im J_T(m_1,m_2;\omega+i\epsilon,\modp) &=& \pi\qint \frac{1}{4E_1 E_2}\left\{\left[1+n_B(E_1)+n_B(E_2)\right]  \left[\delta(\omega-E_1-E_2)-\delta(\omega+E_1+E_2)\right] \right.\nonumber\\ &+&\left. \left[n_B(E_1)-n_B(E_2)\right]  \left[\delta(\omega+E_1-E_2)-\delta(\omega-E_1+E_2)\right] \right\}
\label{imJ}
\end{eqnarray}
where ${\cal P}$ denotes Cauchy's principal value. Note that $\re J$ is even in $\omega$ whereas $\im J$ is odd in $\omega$ as it corresponds to a spectral function.

The $T=0$ part of the above functions corresponds to take all $n_B$ functions as vanishing and is equal to $J(s)$ in the notation of \cite{Gasser:1984gg}, with $s=\omega^2-\modp^2$. The explicit expression for $T=0$ is given in that paper and we do not reproduce it here. The DR UV pole proportional to $\lambda$ in (\ref{lambdadr}) is contained in $J(s=0)=-2\lambda +$ finite terms.

In the general $T\neq 0$ case, the $J_T$ function depends  on $\omega$ and $\modp$ separately due to the breaking of Lorentz covariance in the heat bath. For the case of equal masses $m_1=m_2$, $J_T$ reduces to the $J_0$ function analyzed in \cite{GNLEP02} for thermal pion scattering.

\begin{figure}
\includegraphics[scale=0.5]{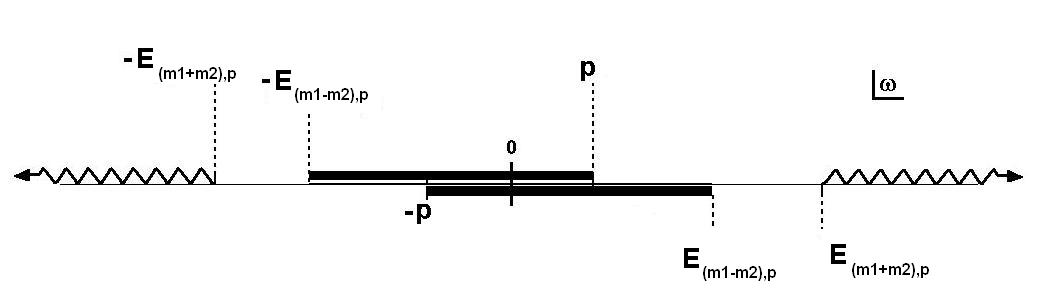}
\caption{Cut structure of the loop integral $J_T(m_1,m_2;\omega,\modp)$ in the $\omega$ complex plane with $E_{(m_1\pm m_2),p}\equiv\sqrt{\modp^2+(m_1\pm m_2)^2}$ and $p\equiv\modp$.}
\label{fig:cutres}
\end{figure}

The imaginary part in (\ref{imJ}) is nonzero along the cuts depicted in Fig.\ref{fig:cutres} in the $\omega$ complex plane. A detailed account of the contributions to the imaginary part for every cut can be found for instance in \cite{Ghosh:2009bt}.  The $\delta(\omega-E_1-E_2)$ and $\delta(\omega+E_1+E_2)$ terms in (\ref{imJ}) require $\omega^2\geq \modp^2+(m_1+m_2)^2$ to be nonzero, for $\omega>0$ and $\omega<0$ respectively. Those two terms account  physically for the decay of a particle $P$ with energy and momentum $(\omega,\vec{p})$ into a pair $P\rightarrow 12$ and the inverse process $12\rightarrow P$, or equivalently to the direct and inverse scattering processes with intermediate states 12 and $s=\omega^2-\modp^2$ the Mandelstam variable. Therefore, this is the usual $T=0$ cut giving rise to unitarity, where the factor $n_B(E_1)+n_B(E_2)$ enhance the contribution of the imaginary part due to the presence of 1 and 2 particles in the thermal bath. On the other hand, the terms proportional to $n_B(E_1)-n_B(E_2)$ give rise to the so called Landau cuts, which are purely thermal, and require  $\omega^2\leq \modp^2+(m_1-m_2)^2$. These Landau cuts arise from processes like $1\rightarrow P2$ and $2\rightarrow P1$ from thermally distributed states $1$ and $2$.  Thus, the $\delta(\omega-E_1+E_2)$ term produces two contributions, one for $\omega\geq\sqrt{(m_1+m_2)^2+\modp^2}$ and another one for $-\modp\leq\omega\leq \modp$, which are depicted as two overlapping cuts in Fig.\ref{fig:cutres}. The same happens for the $\delta(\omega-E_1+E_2)$ term, giving rise to the remaining cuts.

An important  case for this paper is $m_1=0$, $m_2=M$, $\omega^2=\modp^2+M^2$ (on-shell) for which we find, from (\ref{reJ}):

\begin{equation}
\re J_T(0,M;\modp)=\re J_{T=0}(0,M) + \frac{1}{16\pi^2}\frac{1}{\modp}{\cal P}\int_0^\infty dq q \frac{n_B(E_q)}{E_q} \log\left(\frac{\modp+q}{\modp-q}\right)^2
\label{rejtphex}
\end{equation}

where the $T=0$ part can be obtained from the expressions for $J$ in  \cite{Gasser:1984gg} and reads:

\begin{equation}
\re J_{T=0}(0,M)=-2\lambda+\frac{1}{16\pi^2} \left(1-\log\frac{M^2}{\mu_\chi^2}\right)
\label{rejtphexT0}
\end{equation}

Note that in passing from (\ref{reJ}) to (\ref{rejtphex}), the $n_B(E_1)=n_B(q)$ term, which contained an integrable singularity at $q=0$, vanishes exactly and in the $n_B(E_2)$ term, the change of variable $\vec{q}\rightarrow \vec{q}+\vec{p}$ has been performed, so that the integrable singularity at $q=0$ moves to $q=\modp$.

A particularly interesting limit is the static one $\vec{p}=\vec{0}$. Taking this limit in our previous expression  (\ref{rejtphex}) yields:

\begin{equation}
\re J_T(0,M;\vert\vec{p}\vert\rightarrow 0^+)=\re J_{T=0}(0,M)+g_2(M,T)
\label{rejtphexp0}
\end{equation}
with:

\begin{equation}
g_2(M,T)=\frac{1}{4\pi^2}\int_0^\infty dq \frac{n_B(E_q)}{E_q}=-\frac{d g_1(M,T)}{M^2}
\label{g2}
\end{equation}
which behaves asymptotically as $g_2(M,T)\simeq\frac{T}{8\pi M}$ for $T\gg M$ and $g_2(M,T)\simeq (1/2M^2)(2\pi)^{-3/2}(M/T)^{3/2}e^{-M/T}$ for $T\ll M$.

The analysis of the imaginary part for the case of one vanishing mass and on-shell external line is relevant for our discussion in section \ref{sec:im}. In this case, the Landau and unitarity cuts in Fig.\ref{fig:cutres} meet at the branch points $\omega^2=\vert\vec{p}\vert^2+M^2$ ($\omega=\pm E_p$) i.e. precisely at the physical on-shell point. Starting from the general expression (\ref{imJ}), the first $\delta$ function requires in that case $E_p=q+\sqrt{\vert\vec{q}-\vec{p}\vert^2+M^2}$. That condition holds only for $q=0$, provided $M>0$ (so that the other solution at $\frac{\vec{p}\cdot\vec{q}}{\modp q}\equiv\cos\theta=E_p/\vert\vec{p}\vert>1$ is discarded). Hence, $\delta(E_p-q-E_2)=\frac{E_p\delta(q)}{E_p-\vert\vec{p}\vert\cos\theta}$ so that the angular integration in $\theta$ can be easily performed, and so on for $\omega=-E_p$ in the second $\delta$ contribution in (\ref{imJ}). The third  and fourth $\delta$ contribution for this case require also $q=0$, with $\omega=E_p$ for the third one and $\omega=-E_p$ for the fourth. Now, because of the $\delta(q)$, in all these terms the only surviving contributions are those proportional to $n_B(q)$, for which the integrand behaves near $q\rightarrow 0^+$ as $q^2 \frac{n_B(q)}{q} \sim T$. In particular, the $T=0$ contribution vanishes, as it corresponds to the absence of  bremsstrahlung for a charged scalar particle in vacuum. In addition, we should take into account that our $\delta$-functions come from the separation in (\ref{Jcomplex}) $\frac{1}{x+i\epsilon}={\cal P}\frac{1}{x}-i\pi\delta(x)$ so that $\delta(x)=\frac{1}{\pi}\frac{\epsilon}{\epsilon^2+x^2}$ and therefore:

 $$\int_0^\infty \delta(x)=\lim_{\epsilon\rightarrow 0^+}\int_0^\infty \frac{1}{\pi}\frac{\epsilon}{\epsilon^2+x^2}=\lim_{\epsilon\rightarrow 0^+} \frac{1}{\pi} \left.\arctan\left(\frac{x}{\epsilon}\right)\right\vert_0^\infty=\frac{1}{2}$$

 Altogether, we find:

\begin{equation}
\im J_T(0,M;\omega=E_p,\vert\vec{p}\vert)=\frac{1}{16\pi}\frac{T}{p}\log\left(\frac{E_p+\vert\vec{p}\vert}{E_p-\vert\vec{p}\vert}\right)
\label{imjtphex}
\end{equation}
which in the $\vert\vec{p}\vert \rightarrow 0^+$ limit becomes $\im J_T(0,M;\vert\vec{p}\vert \rightarrow 0^+)=\frac{T}{8\pi M}$.

An alternative way to arrive to the result (\ref{imjtphex}) is to calculate $\im J_T(0,M;\omega+i\epsilon,\modp)$ for arbitrary $\omega$ off the on-shell point. Taking then the limit $\omega\rightarrow E_p^+$ one can then check that (\ref{imjtphex}) is recovered.

\end{document}